\documentclass[pra,twocolumn]{revtex4-1} % for arxiv
\usepackage{amssymb,amsmath,bm} 
\usepackage{mathrsfs}
\usepackage{cases}
\usepackage{graphicx}
\usepackage{subfigure}
\usepackage{here}

\makeatletter
\def \v#1{{\bm #1}}
\def \be {\begin{equation}}
\def \ee {\end{equation}}
\newcommand{\Exp}[1]{\,\mathrm{e}^{\mbox{\footnotesize$#1$}}}
\newcommand{\I}{\mathrm{i}}

\newcommand{\tr}[1]{\mathrm{tr}\left(#1\right)}
\newcommand{\Tr}[1]{\mathrm{Tr}\left\{#1\right\}}

\newcommand{\Det}[1]{\mathrm{Det}\left\{#1\right\}}

\def \ds{\displaystyle}

\newcommand{\sldin}[2]{\langle#1,#2\rangle_{\rho_{\theta}}}

\def \del{\partial}

\def \cB{{\cal B}}
\def \cE{{\cal E}}

\def \cM{{\cal M}}
\def \cH{{\cal H}}
\def \cX{{\cal X}}

\def \bbr{{\mathbb R}}
\def \bbc{{\mathbb C}}

\def \sofc2{{\cal S}({\mathbb C}^2)}

\newtheorem{theorem}{Theorem}[section]
\newtheorem{lemma}[theorem]{Lemma}

\newtheorem{corollary}[theorem]{Corollary}

\newenvironment{definition}[1][Definition]{\begin{trivlist}
\item[\hskip \labelsep {\bfseries #1}]}{\end{trivlist}}

\newcommand{\qed}{\nobreak \ifvmode \relax \else
      \ifdim\lastskip<1.5em \hskip-\lastskip
      \hskip1.5em plus0em minus0.5em \fi \nobreak
      \vrule height0.75em width0.5em depth0.25em\fi}

\begin{document}

\title{Nuisance parameter problem in quantum estimation theory:\\ 
General formulation and qubit examples}
\author{Jun Suzuki}
\date{\today}
\email{junsuzuki@uec.ac.jp}
\affiliation{
Graduate School of Informatics and Engineering, The University of Electro-Communications,\\
1-5-1 Chofugaoka, Chofu-shi, Tokyo, 182-8585 Japan
}

\begin{abstract}
In this paper, we analyze quantum-state estimation problems when some of the parameters are of no interest to be estimated. 
In classical statistics, these irrelevant parameters are called nuisance parameters 
and this problem is of great importance in many practical applications of statistics. 
However, little is known regarding the effects of nuisance parameters in quantum-state estimation problems. 
The main contribution of this paper is first to formulate the nuisance parameter problem for the quantum-state estimation theory, 
then to propose a method of how to eliminate the nuisance parameters to obtain 
an estimation error bound for the parameters of interest. We also develop useful methods of dealing with 
the nuisance parameter problem in the quantum case and reveal the significant difference from the classical case. 
In particular, we clarify an intrinsic tradeoff relation to estimate the nuisance parameters and parameters of interest. 
The general qubit model is examined in detail to emphasize that we cannot ignore 
the effects of the nuisance parameters in general. 
Several examples in qubit systems are worked out to illustrate our findings. 
\end{abstract}

%\keywords{Quantum-state estimation, nuisance parameter, tradeoff relation}

\maketitle 
%=====================================================================
\section{Introduction}\label{sec:Intro}
This paper addresses an open problem in quantum-state estimation, the nuisance parameter problem, 
which is the common issue when dealing with the state-estimation problem in practice. 
Consider a family of quantum states specified by the large number of unknown parameters. 
In many quantum information processing task, one is not interested 
in estimating all the parameters but only in a certain subset of parameters. 
In classical statistics, this subset of relevant parameters are called {\it parameters of interest}, 
whereas the remaining irrelevant parameters appearing in a parametric model are called {\it nuisance parameters}. 
This nuisance parameter problem is of great importance in many practical applications of statistics as was formulated by Fisher \cite{fisher35}. 
In the classical case, it is known how much estimation errors become worse in the presence of nuisance parameters. 
Further, many studies were devoted to find effective construction of good estimators. 
See textbooks \cite{lc,bnc,ANbook} and classical results in Refs.~\cite{basu77,rc87,ak88,bs94,zr94}. 

The nuisance parameter problem in the quantum system has just gotten great attention in the quantum information community. 
This is partly because of advances in quantum metrology in a noisy environment. 
See, for example, Ref.~\cite{glm11,emfd11,ddkg12,ta14,ddjk15,ps14}, 
where many studies found that imprecise knowledge about environments may ruin 
advantages of quantum mechanically enhanced precision measurements. 
Another motivation is more general multi-parameter quantum metrology \cite{sbd16,drc17}. 
In both cases, one faces problems of estimating a certain subset of parameters 
in the presence of noise parameters, which should be treated as nuisance parameters. 

We note in passing that several authors analyzed the nuisance parameter problem 
for specific examples \cite{cdbw14,vdgjkkdbw14,js15,kdr17,ych17} and discussed achievability of bounds. 
However, most authors simply treated the problem as the multi-parameter problem. 
For example, Ref.~\cite{cdbw14} discussed a tradeoff relation 
between two kinds of parameters for a specific noise model; the phase and diffusion constant 
without mentioning this is due to the effect of nuisance parameters. 
Later, their theoretical result was demonstrated in Ref.~\cite{vdgjkkdbw14}. 
One of the key results in the nuisance parameter problem was due to Yang {\it et al} \cite{YCH18}. 
The authors made a solid foundation on the asymptotically achievable bound 
for the infinite sample size limit based on the Holevo bound. 
However, this problem also needs to be formulated for the finite sample size case 
to apply any realistic experimental setting. This is one of the primally motivations to this work. 

Here we summarize the main contribution of this paper. 
We first establish a general formulation for the problem of nuisance parameters in the quantum estimation theory. 
To clarify the meaning of achievability of estimation error bounds, 
we introduce an important class of estimators, called a {\it locally unbiasedness about the parameters of interest}. 
We propose a method of how to eliminate the nuisance parameters from a given estimation error bound 
and this general result is illustrated in several qubit examples. 
We compare several different classes of estimators and bounds to be used 
and find that the effect of nuisance parameters in general cannot be ignored for the finite sample size case. 
Based on our approach, we can clearly argue when we can ignore the effect of nuisance parameters 
and in what sense the bound is achievable. 
%As a byproduct, we also extend the method of parameter orthogonalization in statistics 
%to the quantum case. This method will be useful to study the parameter estimation problem 
%in the presence of nuisance parameters. 

As a specific example of the nuisance parameter problem, 
let us consider a familiar qubit-state parametrized by the standard Stokes parameters. 
Suppose that one is only interested in knowing the expectation value of $\sigma_x$, but not the other two parameters. 
What is then the best measurement and estimator for this task? There is no physicist who 
objects to use the projection measurement along $x$ axis and returns the sample mean as an estimator. 
Why is this optimal or we ask what limits us from using other measurements to do better than this? 
Putting differently, optimality is within which class of measurements and estimators?  
By formulating this simple problem as the nuisance parameter problem, 
we can give an affirmative answer to this question from the statistical point of view.  

Next example is relevant to the problem in quantum metrology. 
A typical example is to estimate the value of phase accompanied by some unitary transformation 
when the system undergoes unavoidable quantum noise. 
This class of estimation problem is of great importance for realizing quantum metrological enhancement 
in laboratory where unknown noises from environment are present. 
In the absence of these noises, the {\it ultimate} precision bound for the one-parameter estimation metrology 
is set by the symmetric logarithmic derivative (SLD) quantum Fisher information, which overcomes the classical shot noise limit. 
Now, what is important is to investigate what happens to the precision bound when some of noise parameters are not completely known. 
It many previous analyses on quantum metrology with quantum noise, 
one calculates the SLD Fisher information about variation with respect only to the parameter of interest. 
Although this bound is commonly regarded as the ultimate one even in the presence of nuisance parameter(s), 
this is not proven when some of noise parameters are not known precisely. 
We show that this is indeed not the case in general and derive a sufficient condition that guarantees such a common conclusion. 
A realistic noise model analyzed in this paper also shows that the effects of nuisance parameters cannot be omitted. 

%%%%%%%%%%%%
The outline of this paper is as follows. 
In Sec.~\ref{sec:Cnui}, we provide a brief summary of some of useful results on the nuisance parameter problem in classical statistics.  
In Sec.~\ref{sec:Qnui}, we first formulate the nuisance parameter problem in the quantum case, 
followed by a proposal of the elimination method in Sec.~\ref{sec:Qnui3} and discussions in Sec.~\ref{sec:Qnui4}.   
%In Sec.~\ref{sec:Qpo}, we extend the parameter orthogonalization method to arbitrary quantum system. 
In Sec.~\ref{sec:1para}, we analyze an important class of quantum-state estimation problems, 
estimating a single parameter in the presence of nuisance parameter(s). 
In Sec.~\ref{sec:Qbit} we analyze the nuisance parameter problem for all possible qubit models 
and provide bounds for these models. 
In Sec.~\ref{sec:Ex}, several examples are worked out to illustrate our general formalism. 
The last section \ref{sec:Conc} summarizes our result and state some of open problems on this topics. 
Appendix \ref{sec:AppQFI} gives the quantum score function and quantum Fisher information. 
%Additional materials on classical statistics are provided in Appendix \ref{sec:AppCstat1}. 
Proofs are postponed in Appendix \ref{sec:AppPr}. 
Appendix \ref{sec:AppSuppC} proves that the proposed method recovers the result in classical statistics. 

%%%%%%%%%%%%%%%%%%%%%%%%%%%%%%%%%%%%%%%%%%
\section{Nuisance parameter problem in classical statistics}\label{sec:Cnui}
This section briefly summarizes the nuisance parameter problem in classical statistics. 
More details can be found in books \cite{lc,bnc,ANbook} 
and relevant papers for this work \cite{basu77,rc87,ak88,bs94,zr94}.

\subsection{Cram\'er-Rao inequality in the presence of nuisance parameters}
Consider an $n$-parameter family of probability distributions $p_{\theta}(x)$ on a real-valued set $\cX$, 
where the $n$-dimensional real parameter $\theta=(\theta_1,\theta_2,\dots,\theta_n)$ 
takes values in an open subset of $n$-dimensional Euclidean space $\Theta\subset\bbr^n$. 
The family $\cM_n:=\{p_\theta\,|\,\theta\in\Theta\}$ is called a {\it statistical model} or a {\it model} in short. 
Throughout the paper, we only concern regular models, meaning that 
it behaves well to avoid any mathematical difficulties. 
%Two of relevant regularity conditions 
%among others are; the mapping $\theta\mapsto p_\theta$ is one-to-one and smoothness of parameter dependence on the model 
%so that we can differentiate $p_\theta$ sufficiently many times. 
%The variations about different parameters are to be linearly independent such that the Fisher information matrix is not singular. 
%More technical regularity conditions can found in the standard textbook \cite{lc}. 

Suppose we are interested in estimating the values of a certain subset of parameters 
$\theta_{I}=(\theta_1,\theta_2,\dots,\theta_m)$ ($m<n$), whereas the remaining set of $n-m$ parameters 
$\theta_{N}=(\theta_{m+1},\theta_{m+2},\dots,\theta_n)$ are of no interest. 
%This kind of situation often occurs in various statistical inference problems 
%and is of great importance for practical applications of statistics.  
We denote this partition as $\theta=(\theta_I,\theta_N)$ and assume the similar partition 
for the parameter set $\Theta=\Theta_I\times\Theta_N$. 
The first group of the parameters $\theta_{I}$ is called the {\it parameter of interest} 
and those of no interest are referred to as the {\it nuisance parameter} in statistics. %\cite{com_nui}. 
%The standard problem of classical statistics is 
%to find a good estimator that minimizes a given cost (risk) function under a certain condition. 
%Here, an estimator returns a parameter value $\theta_I$ when given a sequence of the observed data 
%$x^N=x_1x_2\dots x_N$, which is drawn according to a independently and identically distributed 
%source $p^{(N)}_{\theta}(x^N)=\Pi_{t=1}^N p_{\theta}(x_t)$. Mathematically, 
%it is a map from $\cX^N$ to $\Theta_I$. 
Let $\hat{\theta}_I=(\hat{\theta}_1,\hat{\theta}_2,\dots,\hat{\theta}_m)$ be an estimator 
for the parameters of interest, 
and define the mean square error (MSE) matrix by 
\begin{align} \nonumber
V_{\theta}[\hat{\theta}_I]
&=\left[ \sum_{x\in\cX} p_{\theta}(x)(\hat{\theta}_i(x)-\theta_i)(\hat{\theta}_j(x)-\theta_j)  \right]\\
&=\left[ E_\theta\big[(\hat{\theta}_i(X)-\theta_i)(\hat{\theta}_j(X)-\theta_j) \big] \right]. 
\end{align}
where the matrix index takes values in the index set of parameter of interest, i.e., $i,j\in\{1,2,\dots,m\}$, 
and $E_\theta[f(X)]$ denotes the expectation value of a random variable $f(X)$ with respect to $p_\theta(x)$. 
%By definition, the MSE matrix is an $m\times m$ real and positive semidefinite. 

The fundamental question is to find a precision bound for the parameter of interest. 
There are two different scenarios: one is when the nuisance parameters $\theta_N$ 
are completely known and the other is when $\theta_N$ is not known completely. 
%Importantly, the former is an $m$-parameter problem whose model is 
%$\cM_m=\{p_{\theta_I}|\theta_I\in\Theta_I\subset\bbr^m\}$, and the latter is an $n$-parameter problem; $\cM_n$. 
The well-established result in classical statistics proves the following Cram\'er-Rao (CR) inequality: 
\begin{numcases}{V_{\theta_I}[{\hat{\theta}_I}]\ge }
 \ds(J_{\theta_I\theta_I})^{-1}&\hspace{-12pt} ($\theta_N$ is known) \label{ccrineq1} \\[1ex]
 \ds J^{\theta_I\theta_I}& \hspace{-12pt} ($\theta_N$ is not known)\label{ccrineq2}
\end{numcases}
for any (locally) unbiased estimators ${\hat{\theta}}_I$ about the parameters of interest \cite{com_lu}. 
In this formula, two matrices $J_{\theta_I\theta_I}$ and $J^{\theta_I\theta_I}$ 
are defined by the block matrices of the Fisher information matrix $J_\theta$ 
according to the partition $\theta=(\theta_I,\theta_N)$;
\be\label{fisherblock}
J_\theta=\left(\begin{array}{cc}J_{\theta_I\theta_I} & J_{\theta_I\theta_N} \\[0.1ex] 
J_{\theta_N\theta_I}& J_{\theta_N\theta_N}\end{array}\right), \ 
J_\theta^{-1}=\left(\begin{array}{cc}J^{\theta_I\theta_I} & J^{\theta_I\theta_N} \\[0.1ex] 
J^{\theta_N\theta_I}& J^{\theta_N\theta_N}\end{array}\right),  
\ee
where the $(i,j)$ component of the Fisher information matrix is defined by
\be
J_{\theta,ij}=\sum_{x\in\cX}  p_\theta(x) [\frac{\del }{\del\theta_i}\log p_\theta(x)] [\frac{\del }{\del\theta_j} \log p_\theta(x)]. 
\ee
The inverse of the sub-block matrix in inequality \eqref{ccrineq2}, 
\be\label{cpFI}
(J^{\theta_I\theta_I})^{-1}=J_{\theta_I\theta_I}-J_{\theta_I\theta_N}J_{\theta_N\theta_N}^{-1}J_{\theta_N\theta_I}=:J_\theta(\theta_I|\theta_N),
\ee
is known as the {\it partial Fisher information matrix} about $\theta_I$, 
and it accounts the amount of information that can be extracted about $\theta_I$, which can be estimated from a given date. 
Equality \eqref{cpFI} follows from well-known Schur's complements in matrix analysis. 

%%%%
We remark that the nuisance parameters here are treated as non-random variables. 
When they are random as in the Bayesian setting, 
the above CR inequality in the presence of the nuisance parameters needs to be 
replaced by the Bayesian version, see for example Ref.~\cite{rm97}. 

A simple proof of inequalities \eqref{ccrineq1} and \eqref{ccrineq2} is as follows. 
When $\theta_N$ is known, the model $\cM_n$ is reduced 
to an $m$-dimensional model. %$\cM_m$ without any nuisance parameter. 
Hence, we can apply the standard CR inequality to get inequality \eqref{ccrineq1}. 
When $\theta_N$ is not completely known, on the other hand, %the model is $n$-dimensional.  
consider an estimator ${\hat{\theta}}=(\hat{\theta}_I,\hat{\theta}_N)$ for the all parameters $\theta=(\theta_I,\theta_N)$ 
and denote its MSE matrix by $V_\theta[{\hat{\theta}}]$.  
Then, the CR inequality for this $n$-parameter model is 
\be\label{crineqfull}
V_\theta[{\hat{\theta}}]\ge  J_\theta^{-1}, 
\ee
for any locally unbiased estimator at $\theta$. %${\hat{\theta}}=(\hat{\theta}_I,\hat{\theta}_N):\ \cX^n\rightarrow \Theta=\Theta_I\times\Theta_N$. 
Let us decompose the MSE matrix as 
\be\label{Cmseblock}
V_\theta[{\hat{\theta}}]=\left(\begin{array}{cc}V_{\theta_I} & V_{\theta_I,\theta_N} \\[0.2ex] 
V_{\theta_N,\theta_I}& V_{\theta_N}\end{array}\right), 
\ee
then, applying the projection onto the subspace $\theta_I$ gives the desired result \eqref{ccrineq2}. 
%Some subtlety regarding unbiasedness condition in this proof is discussed in Appendix \ref{sec:AppCstat1}. 

\subsection{Discussions on the classical CR inequality and useful concepts}
We discuss the above result concerning the CR inequality \eqref{ccrineq1} and \eqref{ccrineq2} briefly. 
First, it is important to emphasize that two different scenarios 
deal with two different statistical models. In the presence of nuisance parameter(s), 
what we can do best is to estimate all parameters and hence the precision 
bound is set by the standard CR inequality for the $n$-parameter model. 

%Second, when there exist nuisance parameters, the precision bound 
%$J^{\theta_I\theta_I}$ still depends on the unknown values of $\theta_N$. 
%It is then necessary to eliminate the nuisance parameter from this expression. 
%There are several strategies known in classical statistics (See for example, Ref.~\cite{basu77}).  
%The simplest one is to marginalize the effect of nuisance parameter by 
%taking expectation value of $J^{\theta_I\theta_I}$ with respect to 
%some prior distribution for the nuisance parameter $\theta_N$. 
%The other is to adopt the worst case by calculating 
%$\max_{\theta_N\in\Theta_N}J^{\theta_I\theta_I}$. 
%
%Third, the existence of a sequence of estimators attaining the 
%equality in the asymptotic limit follows from the standard argument. 
%When no nuisance parameter exists, the maximum likelihood estimator (MLE) 
%for the parameter of interest $\theta_I$ saturates the bound. 
%If we have some nuisance parameter in the model, we can 
%also apply the MLE for the all parameter $\theta=(\theta_I,\theta_N)$. 
%This asymptotically saturates the CR inequality \eqref{crineqfull}, 
%so as inequality \eqref{ccrineq2}. 

Second, we have the (asymptotically) achievable precision bound for the MSE matrix given by Eq.~\eqref{ccrineq2}, 
but this bound is not practically useful in general. This is because one has to 
estimate {\it all parameters} in order to achieve it asymptotically by the use of MLE. 
This is usually very expensive to solve the likelihood equation in general. 
In particular, this is the case when the number of nuisance parameters are large compared to that of parameters of interest. 
Thus, there remain practical problems left to find efficient estimators in the presence of nuisance parameters. 
For example, Ref.~\cite{basu77} lists ten different methods of dealing with this problem. 

%Fifth, there exist several different derivations of the CR inequality \eqref{ccrineq2} in the presence of nuisance parameters. 
%Based on each different proof, we can give different interpretations of this result.  
%In Appendix \ref{sec:AppCstat2}, we give two alternative proofs. A nontrivial part of this fact is that 
%all three different methods at first sight lead to the same precision bound. 

Last, it is well known that the following matrix inequality holds. 
\begin{align}\nonumber
J^{\theta_I\theta_I}&=(J_{\theta_I\theta_I}-J_{\theta_I\theta_N}J_{\theta_N\theta_N}^{-1}J_{\theta_N\theta_I})^{-1}\\
&\ge ( J_{\theta_I\theta_I})^{-1}. \label{nuiineq}
\end{align}
Here the equality holds if and only if the off-diagonal block matrix vanishes, i.e., $J_{\theta_I\theta_N} =0$.  
When $J_{\theta_I\theta_N} =0$ holds at $\theta$, we say that two sets of parameters $\theta_I$ and $\theta_N$ are {\it orthogonal with respect 
to the Fisher information} at $\theta$ or simply $\theta_I$ and $\theta_N$ are orthogonal at $\theta$. 
When $J_{\theta_I\theta_N} =0$ holds for all $\theta \in\Theta$, $\theta_I$ and $\theta_N$ are 
said {\it globally orthogonal}. 
%In the next subsection, we discuss some of the consequences of parameter orthogonality. 

Summarizing the result known in classical statistics, 
the MSE becomes worse in the presence of nuisance parameters when compared 
with the case of no nuisance parameters. We can regard the difference of two bounds 
as the loss of information due to the presence of nuisance parameters. 
This quantity is defined by 
\be \label{losscinfo}
\Delta J_\theta^{-1}:= J^{\theta_I\theta_I}- ( J_{\theta_I\theta_I})^{-1}. 
\ee
When the values of $\Delta J_\theta^{-1}$ is large (in the sense of matrix inequality), 
the effect of nuisance parameters is more important. From the above mathematical 
fact, we have that no loss of information is possible if and only if two sets of 
parameters are globally orthogonal, i.e.,
\be
\Delta J_\theta^{-1}=0 \ \Leftrightarrow\ J_{\theta_I\theta_N} =0, 
\ee
for all values of $\theta=(\theta_I,\theta_N)\in\Theta$. 

From the above discussion, the orthogonality condition is a key ingredient 
when discussing parameter estimation problems with nuisance parameters. 
This was pointed out in the seminal paper by Cox and Reid \cite{rc87}. 

%Denote the $i$th partial derivative of the logarithmic likelihood function $\ell_\theta(x):=\log p_\theta(x)$, 
%which is known as the {\it score function}, by 
%\be\label{cscorefn}
%u_{\theta,i}(x):=\frac{\del}{\del\theta_i}\ell_\theta(x).
%\ee 
%Here after, we set the sample size $N=1$ to simplify notation. 
%Then, the $i,j$ component of the Fisher information matrix is 
%calculated by the expectation value of the multiplication as 
%$J_{ij}(\theta)=E_\theta[u_{\theta,i}(X)u_{\theta,j}(X)]$. 
%The orthogonality condition is equivalent to say that two sets of random variables 
%$u_{\theta,I} (X)=(u_{\theta,1} (X),\dots,u_{\theta,m} (X))$ and 
%$u_{\theta,N} (X)=(u_{\theta,m+1} (X),\dots,u_{\theta,n} (X))$ 
%are statistically independent. As an example, consider a two-parameter model 
%with $\theta_2$ a nuisance parameter. 
%When $\theta_1$ and $\theta_2$ are orthogonal, two maximum likelihood estimators (MLEs) 
%$\hat\theta_1$ and $\hat{\theta}_2$ are asymptotically independent. 
%As a consequence of it, asymptotic error for $\theta_1$ becomes same regardless of knowing the true value of $\theta_2$ or not. 
%A familiar example of this phenomenon is when estimating the mean value of a normal distribution 
%without knowing its variance \cite{lc,bnc,ANbook}. 

It is well known that any two sets of parameters can be made orthogonal at each point locally 
by an appropriate invertible map from a given parameterization to the new parameters as 
\be\label{nui_change}
\theta=(\theta_I,\theta_N)\mapsto\xi=(\xi_I,\xi_N)\mbox{ s.t. }\xi_I=\theta_I. 
\ee 
The condition $\xi_I=(\xi_1,\xi_2,\dots,\xi_m)=\theta_I$ ensures that 
the parameters of interest are unchanged while the nuisance parameters can be changed arbitrary. 
%This statement holds for an arbitrary model with any number of parameters \cite{amari}. 
%For example, consider the following new parametrization for the nuisance parameters $\theta_N\mapsto\xi_N$: 
%\begin{align} \label{localortho}
%\xi_I&=\theta_I,\\
%\xi_N&=\theta_N+J^{\theta_N\theta_N}J_{\theta_N\theta_I}\big|_{\theta_I=\theta_I(0)}(\theta_I-\theta_I(0)), 
%\end{align}
%where $\theta_I(0)=(\theta_1(0),\dots,\theta_m(0))$ is an arbitrary reference point. 
However, the global orthogonalization of parameters is in general impossible unless the model satisfies 
the special conditions. One such instance is when the number of parameter of interest is one ($m=1$), 
and the rest of parameters is the nuisance parameter, that is, 
$\theta_I=\theta_1, \theta_N=(\theta_2,\dots,\theta_n)$ \cite{rc87}. 
This Cox-Reid method can be extended to the quantum case straightforwardly. 
Its exposition of this method in the quantum case will be presented elsewhere. 

\section{Formulation of the problem}\label{sec:Qnui}
\subsection{Quantum parameter estimation problem}\label{sec:Qnui1}
In this subsection, we shall briefly summarize the result of 
the quantum estimation theory \cite{helstrom,holevo,ANbook,hayashi,petz} for more details. 
%Ref.~\cite{js15} also provides an introductory and concise summary 
%for the parameter estimation theory in quantum systems, which is directly related to this paper. 

A {\it quantum system} is represented by a $d$-dimensional complex vector space $\cH=\bbc^d$. 
%When the dimension of the system is two, we speak of ``qubit" that is the simplest quantum system. 
To simplify our discussion we only consider quantum systems with a fixed dimension $d<\infty$. 
A {\it quantum state} $\rho$ is a nonnegative matrix on $\cH$ with unit trace. 
%The set of all quantum states on $\cH$ is denoted by $\sofh:=\{\rho\,|\,\rho\ge0,\tr{\rho}=1 \}$.

A {\it measurement} $\Pi$ is a set of nonnegative matrices 
$\Pi=\{\pi_x\}_{x\in\cX}$ such that the condition $\sum_{x\in\cX}\Pi_x=I_d$ ($I_d$: $d\times d$ identity matrix) is satisfied, 
where $\cX$ is a label set for the measurement outcomes. 
The set $\Pi$ is referred to as the positive operator-valued measure (POVM). 
When the POVM consists of mutually orthogonal projectors, we call it the projection valued measure (PVM) 
or simply the projection measurement. 
The probabilistic rule when a POVM $\Pi$ is performed on $\rho$ is given by 
\begin{equation}\label{born}
p_\rho(x|\Pi):=\tr{\rho\Pi_x}, 
\end{equation}
which corresponds to the probability of the measurement outcome $X=x$.  

A {\it quantum statistical model} or simply a model is defined by a parametric family of quantum states $\rho_\theta$ on $\cH$: 
\be
\cM_n:=\{\rho_\theta\,|\,\theta\in\Theta\subset\bbr^n\}. 
\ee
As in the classical statistics, we assume necessary regularity conditions \cite{com_reg}. 
A set of a measurement $\Pi$ and an estimator $\hat\theta$, $\hat{\Pi}=(\Pi,\hat{\theta})$, 
is called a {\it quantum estimator} or simply an {\it estimator}. 
We define the MSE matrix for the estimator $\hat{\Pi}$ by 
\begin{align}\nonumber
V_{\theta}[\hat{\Pi}]
&=\left[ \sum_{x\in\cX} \tr{\rho_\theta\Pi_x}({\hat{\theta}_i}(x)-\theta_i)({\hat{\theta}_j}(x)-\theta_j)  \right]\\
&=\left[ E_\theta\big[({\hat{\theta}_i}(X)-\theta_i)({\hat{\theta}_j}(X)-\theta_j)|\Pi\big]  \right].
\end{align}
where $E_\theta[f(X) |\Pi]$ is the expectation value of a random variable $f(X)$ 
with respect to the distribution $p_{\rho_\theta}(x|\Pi)=\tr{\rho_\theta\Pi_x}$. 
The aim of parameter estimation problems for the quantum case is to 
find an optimal estimator $\hat{\Pi}=(\Pi,\hat{\theta})$ such that the MSE matrix 
is minimized as much as possible. 

We note that it is in general not possible to minimize the MSE matrix over all possible measurements 
in the sense of a matrix inequality. This kind of situation often happens in the theory of optimal design of experiments, 
where one tries to minimize the inverse of the Fisher information matrix over design parameters. 
See the textbooks \cite{fedorov,pukelsheim,fh97,pp13,fl14} and Ref.~\cite{gns19} in the context of quantum estimation theory.  
One of possible approaches is to optimize the weighted trace of the MSE matrix: 
\be
 \Tr{W_nV_{\theta}[\hat{\Pi}]}, 
\ee
for a given positive matrix $W_n$. Here, the matrix $W_n$ is called a {\it weight matrix}. 
%(also called a {\it utility matrix} or {\it loss matrix} in statistical literature)
%and it represents a tradeoff relation upon estimating different vector component of the parameter $\theta$. 
%For instance, the case $W_n=I_n$ (the $n\times n$ identity matrix) corresponds 
%to minimize the averaged variance of estimators. In the language of optimal design of experiments, 
%this optimality is called the A-optimal design. We can similarly define other optimality functions 
%to define optimal estimators \cite{fedorov,pukelsheim,fh97,pp13,fl14}. 
We mainly consider strictly positive weight matrices, i.e., $W_n>0$, 
although it is also possible to formulate the problem with a nonnegative weight matrix. 
However, as we will discuss in this paper, a role of the weight matrix is important 
when discussing the nuisance parameter problem in the quantum case. 

The main interest in this paper is to find the precision bound under the locally unbiasedness condition on estimators. 
An estimator $\hat{\Pi}$ is {\it locally unbiased} at $\theta$ if 
\[
E_\theta[{\hat{\theta}_i}(X)|\Pi]=\theta_i\ \mathrm{and}\  \frac{\del}{\del\theta_j}E_\theta[{\hat{\theta}_i}(X)|\Pi]=\delta_{ij}
%\sum_{x\in\cX} {\hat{\theta}_i}(x) \tr{\rho_{\theta}\Pi_x}=\theta_i, \ 
%\sum_{x\in\cX} {\hat{\theta}_i}(x)\tr{\frac{\del}{\del\theta_j}\rho_{\theta}\Pi_x}=\delta_{ij}, 
\]
are satisfied at $\theta\in\Theta$ for all parameter indices $i,j\in\{1,2,\dots,n\}$. 
%Note that this condition is to require the usual unbiasedness condition at a point $\theta$ up to 
%the first order in the Taylor expansion. 
The fundamental precision bound is defined by 
\be\label{qcrbound}
C_\theta[W_n,\cM_n]:=\min_{\hat{\Pi}\mathrm{\,:l.u.at\,}\theta}\Tr{W_nV_\theta[\hat{\Pi}]}, 
\ee
where the minimization is carried out for all possible estimators under the locally unbiasedness condition, 
which is indicated by l.u.~at $\theta$ \cite{com_collective}. 
In this paper, any bound for the weighted trace of the MSE matrix is referred to as the {\it CR type bound}. 
When a CR type bound is tight as in Eq.~\eqref{qcrbound}, we call it as the {\it most informative bound} in our discussion.  
Here we list some of familiar examples of CR type and most informative bounds: 
The SLD CR bound for a one-parameter model \cite{helstrom}, 
%the RLD CR bound for a Gaussian shift model \cite{yl73,holevo}, 
the Nagaoka bound for a two-parameter qubit model \cite{nagaoka89}, 
and the Hayashi-Gill-Massar bound for three parameter qubit model \cite{hayashi97,GM00}. 

%Note this bound in general depends on the value of parameter $\theta$ and the choice of the weight matrix $W_n$. 
%Let $\hat{\Pi}_{\mathrm{opt}}:=\arg\min  \Tr{W_nV_\theta[\hat{\Pi}]}$ be an optimal estimator attaining the minimum 
%of the CR type bound \eqref{qcrbound}, then it is clear that this $\hat{\Pi}_{\mathrm{opt}}$ represents 
%the best measurement and the estimator in the sense of the above optimization. 
%That is, if somebody specifies the weight matrix $W_n$, 
%we can always construct the best estimator $\hat{\Pi}_{\mathrm{opt}} $ that minimizes 
%the weighted trace of the MSE. 

%When considering positive semi-definite weight matrices, the CR type bound cannot be 
%attained explicitly in general. In this case, we have
%\be\label{qcrbound2}
%\underline{C}_{\,\theta}[W_n,\cM_n]:=\inf_{\hat{\Pi}\mathrm{\,:l.u.at\,}\theta}\Tr{W_nV_\theta[\hat{\Pi}]}, 
%\ee
%for $W_n\ge0$. 
%The difference from bound \eqref{qcrbound} is that 
%an optimal estimator may not be locally unbiased at $\theta$ for low-rank matrices $W_n$. 

An important remark regarding this ``optimal estimator" is 
that it depends on the unknown parameter value $\theta$ in general. 
%due to the structure of the above optimization problem. 
%In other words, one has to perform these measurements to estimate 
%unknown parameters by using unknown values $\theta$. 
%This contradictory fact creates a major opponent against use of (locally) unbiased estimators in 
%classical statistics. Here, we stress that methods of statistical inference 
%provide an additional ingredient to overcome such a difficulty and to achieve 
%bound \eqref{qcrbound} asymptotically. 
There are several methods to overcome this problem. 
The {\it adaptive state-estimation method}, which was proposed by 
Nagaoka in the context of quantum-state estimation problems \cite{nagaoka89-2}. 
Rigorous treatment of such a scheme is due to Fujiwara \cite{fujiwara06}. 
%We also mention that alternative approaches exist to achieve 
%bound \eqref{qcrbound} such as 
Another well-known method is the two-step estimation method \cite{HM98,BNG00}. 
Other various types of adaptive schemes have been intensively studied recently. 
See, for example, Refs.~\cite{stm12,oioyift12,mrdfbks13,ksrhhk13,hzxlg16,ooyft17} 
and a review paper \cite{zlwjn17} and references therein. 

Before we move onto the discussion on the nuisance parameter problem, 
we show an alternative expression for the most informative bound, which is due to Nagaoka \cite{nagaoka89}. 
%He proved the above bound can alternatively be expressed as the following optimization.  
For a given quantum statistical model $\cM_n=\{\rho_\theta|\theta\in\Theta\}$, 
let us fix a POVM $\Pi$. %=\{\Pi_x\}_{x\in\cX}$. 
Then, the probability distribution determined by measurement outcomes $p_\theta(x|\Pi)=\tr{\rho_\theta\Pi_x}$ 
defines a classical statistical model: 
%\be
$\cM(\Pi):=\{p_\theta(\cdot|\Pi)\,|\,\theta\in\Theta\}$. 
%\ee
%If the resulting classical model is regular, we can calculate 
%the Fisher information matrix $J_\theta[\Pi]$ about this model, and the CR inequality 
%holds for the MSE matrix. Therefore, one can show that 
Then, we have
\be\label{MICRbound}
C_\theta[W_n,\cM_n]=\min_{\Pi\mathrm{: POVM}} \Tr{W_n J_\theta[\Pi]^{-1} }, 
\ee
where the minimum is taken over all possible POVMs. 
%It is important to note that statistical model $\cM(\Pi)$ can violate regularity conditions for some POVM. 
%For example, the Fisher information matrix is singular. In this case, one cannot 
%calculate the inverse directly. A standard treatment is to use the generalized inverse with some care \cite{sm01}. 
%Alternatively, regularization techniques are often used in literature. 
%In the above optimization in Eq.~\eqref{MICRbound}, due to the positivity assumption of the weight matrix, 
%one automatically exclude POVMs giving rise to non-regular statistical models. 
%That is, we minimize the weighted trace of the inverse of Fisher information matrix 
%within regular POVMs meaning that their statistical models are regular (See also Appendix \ref{sec:AppPr3}.).  

\subsection{Nuisance parameter problem in quantum case}\label{sec:Qnui2}
We now introduce a model with nuisance parameters for the quantum case. 
Consider an $n$-parameter model as before and divide the parameters into two groups, 
one consists of parameters of interest $\theta_I=(\theta_1,\theta_2,\dots,\theta_m)$ 
and the other subset is the nuisance parameters $\theta_N=(\theta_{m+1},\theta_{m+2},\dots,\theta_n)$. 
We thus have a family of quantum states parametrized by two different kinds of parameters:
\be
\cM_n=\{\rho_\theta\,|\,\theta=(\theta_I,\theta_N)\in\Theta\}. 
\ee
Our goal is to perform a good measurement and then to infer the values of parameter of interest $\theta_I$. 
Let $\hat{\Pi}_I=(\Pi,\hat{\theta}_I)$ be an estimator for the parameter of interest 
and define its MSE matrix for the parameters of interest by 
\begin{align}\nonumber
V_{\theta_I}[\hat{\Pi}_I]
&=\left[ \sum_{x\in\cX} \tr{\rho_\theta\Pi_x}({\hat{\theta}_i}(x)-\theta_i)({\hat{\theta}_j}(x)-\theta_j)  \right]\\
&=\left[ E_\theta\big[({\hat{\theta}_i}(X)-\theta_i)({\hat{\theta}_j}(X)-\theta_j)|\Pi \big] \right],
\end{align}
where the matrix indices $(i,j)$ run from $1$ to $m$, but not from $1$ to $n$. 
Hence, the MSE matrix is an $m\times m$ matrix. 
We wish to find the precision bound for the above MSE matrix about the parameter of interest 
under the locally unbiasedness condition. 

\subsubsection{Locally unbiasedness condition for the parameter of interest}
When dealing with the nuisance parameter problem, it is necessary 
to define the locally unbiasedness for a subset of parameters. 
%(See also Appendix \ref{sec:AppCstat1}.)
Let us consider the two sets of parameters $\theta=(\theta_I,\theta_N)$ and 
an estimator $\hat{\theta}_I=(\hat{\theta}_1,\dots,\hat{\theta}_m)$ as before. 
An estimator $\hat{\theta}_I$ for the parameter of interest is said {\it unbiased} 
for $\theta_I$ at $\theta$, if the condition 
\be
E_\theta[{\hat{\theta}_i}(X)|\Pi]=\theta_i,
\ee
holds for all $i=1,2,\dots,m$ and for all $\theta\in\Theta$. 
%Clearly, this condition of unbiasedness does not concern about the estimate of the nuisance parameters. 

We next introduce the cencept of locally unbiasedness for the parameter of interest as follows. 
\begin{definition}
An estimator $\hat{\theta}_I$ for the parameter of interest is locally unbiased for $\theta_I$ at $\theta$, 
if, for $\forall i\in \{1,\dots,m\}$ and $\forall j\in \{1,\dots,n\}$, two conditions, 
\be \label{lu_cond}
E_\theta[{\hat{\theta}_i}(X)|\Pi]=\theta_i\ \mathrm{and}\  \frac{\del}{\del\theta_j}E_\theta[{\hat{\theta}_i}(X)|\Pi]=\delta_{ij}
\ee
are satisfied at a given point $\theta$. 
\end{definition}
An important condition here is an additional requirement: 
$\frac{\del}{\del\theta_j}E_\theta[{\hat{\theta}_i}(X)|\Pi]=0$ for $i=1,2,\dots,m$ and $j=m+1,m+2,\dots,n$. 
This condition can be trivially satisfied if a probability distribution from a POVM 
is independent of the nuisance parameters. But this can only happen in special cases. 
In general, non-vanishing of $\frac{\del}{\del\theta_j}E_\theta[{\hat{\theta}_I}(X)|\Pi]$ 
for $j=m+1,m+2,\dots,n$ contributes the estimation error bound for the parameters of interest. 
%see the general inequality \eqref{mse_genineq} in Appendix \ref{sec:AppCstat1}. 

It is known that for a given regular statistical model, we can always construct a 
locally unbiased estimator at arbitrary point (See expression \eqref{lu_est} in Appendix \ref{sec:AppPr2}.). 
We can extend a similar construction for a locally unbiased estimator for the parameter of interest as follows. 
Given a statistical model in which the score functions for the nuisance parameters 
are not linearly independent, i.e., $\{\frac{\del}{\del \theta_i}\log p_\theta(x)\}_{i=m+1,\dots,n}$ are linearly dependent. 
In this case, the Fisher information matrix is singular and is not invertible. 
Even in this case, the following estimator is locally unbiased for $\theta_I=(\theta_1,\dots,\theta_m)$:
\be
\hat{\theta}_i(x)=\theta_i+\sum_{j=1}^m \left( J(\theta_I|\theta_N) \right)^{-1}_{ji} u_{\theta_I,j}(x|M_*),
\ee
where $J(\theta_I|\theta_N)$ is the partial Fisher information of Eq.~\eqref{cpFI}, 
and $u_{\theta_I,j}(x|M_*)=$ are the effective score functions \cite{bs94}, %defined in Appendix \ref{sec:AppCstat2}. 
where $u_{\theta_I,i}(x|M):=\del_i \log p_\theta(x)-\sum_{j=m+1}^nm_{ij}\del_j\log p_\theta(x)$ 
for a given $M=[m_{ij}]$ ($m\times (n-m)$ real matrix) and $M_*=J_{\theta_I\theta_I} (J_{\theta_N\theta_N})^{-1}$. 

At first sight, the above definition \eqref{lu_cond} depends on the nuisance parameters explicitly.  
One might then expect that the concept of locally unbiased estimator for $\theta_I$ 
is not invariant under reparametrization of the nuisance parameters of the form \eqref{nui_change}. 
The following lemma shows the above definition in fact does not depend on parametrization of the nuisance parameters. 
Its proof is given in Appendix \ref{sec:AppPr1}.
\begin{lemma}\label{lem_lucond}
If an estimator $(\Pi,\hat{\theta}_I)$ is locally unbiased for $\theta_I$ at $\theta$, 
then it is also locally unbiased for the new parametrization defined by the transformation \eqref{nui_change}. 
That is, if two conditions \eqref{lu_cond} are satisfied, then the following conditions also hold. 
\be \label{lu_cond2} 
E_\xi[{\hat{\theta}_i}(X)|\Pi]=\xi_i\ \mathrm{and}\  \frac{\del}{\del\xi_j}E_\xi[{\hat{\theta}_i}(X)|\Pi]=\delta_{ij},
\ee
for $\forall i\in \{1,\dots,m\}$ and $\forall j\in \{1,\dots,n\}$. 
\end{lemma}

Having introduced the locally unbiasedness condition for the parameter of interest, 
we define the most informative bound for the parameter of interest by the following optimization: 
\begin{definition}
For a given $m\times m$ weight matrix $W_I>0$, the most informative bound about the parameter of interest is 
defined by
\be\label{qcrboundnui0}
C_{\theta_I}[W_I,\cM_n]:=\min_{\hat{\Pi}_I\mathrm{\,:l.u.\,for\,}\theta_I} \Tr{W_IV_{\theta_I}[\hat{\Pi}_I]}, 
\ee
where the condition for the minimization is such that estimators $\hat{\Pi}_I$ are 
locally unbiased for $\theta_I$ at $\theta$. 
\end{definition}

By extending the argument of deriving the relationship \eqref{MICRbound} 
and the classical CR inequality \eqref{ccrineq2} in the presence of nuisance parameter, 
we can show that the following alternative expression holds (See Appendix \ref{sec:AppPr2} for its derivation.). 
\be\label{MICRbound2}
C_{\theta_I}[W_I,\cM_n]=\min_{\Pi\mathrm{: POVM}} \Tr{W_I J^{\theta_I\theta_I}[\Pi] }, 
\ee
where $J_\theta^{\theta_I\theta_I}[\Pi] $ is the block submatrix of the Fisher information 
matrix about the POVM $\Pi$ (C.f., Eq.~\eqref{fisherblock}).
%It may happen that above direct minimization \eqref{MICRbound2} is 
%harder than the case of estimating all parameters in general. 

For our discussion, it is convenient introduce the following notation and definitions. 
For an $n$-parameter model, we also consider an estimator for the nuisance parameter $\theta_N$ 
and denote it by $\hat{\theta}_N$. We then have a quantum estimator for all parameters 
as $\hat{\Pi}=(\Pi,\hat{\theta}_I,\hat{\theta}_N)$. Let $V_\theta[\hat{\Pi}]$ be 
the $n\times n$ MSE matrix about the estimator $\hat{\Pi}$, and introduce the following 
block matrix decomposition:
\be\nonumber
V_\theta[\hat{\Pi}]=\left(\begin{array}{cc}V_{\theta_I} [\hat{\Pi}]& V_{\theta_I\theta_N}[\hat{\Pi}] \\[0.1ex] 
V_{\theta_N\theta_I}[\hat{\Pi}] & V_{\theta_N}[\hat{\Pi}]\end{array}\right),
\ee
which is same as in the classical case \eqref{Cmseblock}. Likewise, 
for the weight matrix $W_n$, consider the same decomposition: 
\be \nonumber
W_n=\left(\begin{array}{cc}W_{I} & W_{IN} \\[0.1ex] 
W_{NI}& W_{N}\end{array}\right).
\ee

\subsubsection{Class of estimators and possible bounds}
Let us compare two important classes of quantum estimators in our discussions. 
Denote by $\cE_\theta$ a set of all estimators that are locally unbiased about $\theta=(\theta_I,\theta_N)$ 
at $\theta$ and by $\cE_{\theta_I}$ locally unbiased estimators about $\theta_I$ at $\theta$: 
\begin{align}\nonumber
\cE_\theta&:=\{\hat{\Pi}\,|\,\hat{\Pi}\mbox{ is locally unbiased about }\theta\},\\
\cE_{\theta_I}&:=\{\hat{\Pi}\,|\,\hat{\Pi}_I\mbox{ is locally unbiased about }\theta_I\}. 
\end{align} 
Obviously, the set $\cE_{\theta}$ is a subset of $\cE_{\theta_I}$, since we impose 
additional conditions for $\hat{\Pi}$, that is ${\cE}_{\theta}\subset{\cE}_{\theta_I}$ holds.  
We now show two alternative estimation error bounds and discuss their relations to 
the most informative bound for the parameter of interest \eqref{qcrboundnui0}. 

\noindent
\textit{Method 1.} (Locally unbiased estimation method)\\
Having in mind that optimal POVMs depend on the nuisance parameter $\theta_N$ in general, 
we may look for a restricted class of estimators. 
For an estimator $\hat{\Pi}=(\Pi,\hat{\theta}_I,\hat{\theta}_N)$ about $\theta$, 
we restrict an estimator to consider $\hat{\Pi}_I=(\Pi,\hat{\theta}_I)$ for the parameter of interest. 
We impose the locally unbiasedness condition on both parameters $\theta=(\theta_I,\theta_N)$ 
and consider the optimization problem: 
\be\label{qcrboundnui1}
C^{\cE_{\theta}}_{\theta_I}[W_I,\cM_n]=\inf_{\hat{\Pi}\in\cE_{\theta}}\Tr{W_IV_{\theta_I}[\hat{\Pi}]}. 
\ee
The difference from Eq.~\eqref{qcrboundnui0} is that estimators here 
need to satisfy the additional locally unbiasedness condition. 
Furthermore, we may not be able to find an optimal estimator that 
is locally unbiased for all parameters $\theta$. 
This is why the bound is given by the infimum. 
%Due to the inclusion relationship $\cE_{\theta}\subset\cE_{\theta_I}$, 
%we have an inequality $C^{\cE_{\theta}}_{\theta_I}[W_I,\cM_n]\ge C_{\theta_I}[W_I,\cM_n]$ in general. 

\noindent
\textit{Method 2.} (The weight-matrix limit method)\\
We consider an $n$-parameter model $\cM_n$ including nuisance parameter. 
Let $C_{\theta}[W_n,\cM_n]$ be a CR type bound, 
and we take the following limit for the weight matrix $W_n$: 
Letting the $n\times n$ weight matrix to the $m\times m$ matrix by 
\be \label{weightlimit}
W_n\to \lim_{\epsilon\to0^+}
\left(\begin{array}{cc}W_{I} & 0 \\[0.1ex] 
0& \epsilon I_N\end{array}\right)\ =\ \left(\begin{array}{cc}W_{I} & 0 \\[0.1ex] 
0&0\end{array}\right),  
\ee
where $I_N$ denotes the $(n-m)\times(n-m)$ identity matrix. 
There exist several other limiting procedures, but we only consider 
the above case in this paper. 
In this way, we obtain the relevant bound: 
\begin{align}\label{qcrboundnui2}
C^{\lim}_{\theta_I}[W_I,\cM_n]&=\lim_{W_n\to W_I}C_\theta[W_n,\cM_n]\\ \nonumber
&=\lim_{W_n\to W_I}\min_{\hat{\Pi}\in\cE_\theta}\Tr{W_nV_\theta[\hat{\Pi}]}. 
\end{align}
Here the matrix limit indicates procedure \eqref{weightlimit}. 

We now discuss three bounds defined so far: 
$C_{\theta_I}[W_I,\cM_n],C^{\cE_{\theta}}_{\theta_I}[W_I,\cM_n],C^{\lim}_{\theta_I}[W_I,\cM_n]$. 
Note that three bounds are defined based on different estimation strategies. 
The following theorem proves that they are all equivalent. 
The proof is given in Appendix \ref{sec:AppPr3}.  
\begin{theorem}\label{thm_equiv} 
For our setting, $C_{\theta_I}[W_I,\cM_n]=C^{\cE_{\theta}}_{\theta_I}[W_I,\cM_n]=C^{\lim}_{\theta_I}[W_I,\cM_n]$ 
holds for all $W_I>0$. 
\end{theorem}

Let us make a few comments on the nature of these (equivalent) bounds.  
First, it is clear that the direct minimization problem \eqref{qcrboundnui0}, 
or equivalently \eqref{MICRbound2}, is the fundamental bound. 
When the optimal measurement depends on the nuisance parameters, 
it is not obvious how to achieve this bound. 
Additional discussion is needed to clarify achievability.  

Second, regarding method 2. We first note that the procedure of setting the weight matrix 
to the singular matrix needs a special care. Beside some mathematical subtlety in matrix limits, 
we still have a question of achievability for a such derived bound in limit \eqref{weightlimit}. 
This is because it is not obvious the limit of optimal estimators exits or not. 
That is to say, limit \eqref{weightlimit} may results in a not meaningful measurement. 
It may also happen that the resulting POVM depends on the value of nuisance parameter. 
If this is the case, bound \eqref{qcrboundnui2} is not (asymptotically) achievable, 
and it holds only in the sense of infimum. 

Third, the locally unbiasedness condition for the parameter of interest. 
At first thought, this condition is well-defined notion as it is sufficient to prove the classical CR inequality \eqref{ccrineq2}. 
However, in the quantum case, it seems that this condition may not result in a tight bound. 
An obvious reason is that locally unbiased estimators about $\theta_I$ may not 
be sufficient to gain information about the nuisance parameter in general. 
Thus, we cannot completely ignore the effect of nuisance parameters in the quantum case. 

Last, tradeoff relations among estimation errors. 
One of our motivation in this work is to discuss tradeoff relations 
for estimation errors between the parameter of interest and the nuisance parameter. 
If we follow the optimization methods described here to derive a precision bound, we will not be able to this kind of tradeoff relation. 
This is because we only consider the estimation errors for the parameter of interest in these methods. 
See also discussion in Sec.~\ref{sec:Qnui4}. 

\subsection{Elimination of nuisance parameters in quantum-state estimation problems}\label{sec:Qnui3}
Given an $n$-parameter model $\cM_n$, we consider an estimation problem 
for all parameters under a certain condition, for example, 
the locally unbiasedness condition for both $\theta=(\theta_I,\theta_N)$. 
Let $\hat{\Pi}=(\Pi,\hat{\theta})$ be an estimator for $\theta=(\theta_I,\theta_N)$ 
and denote its MSE matrix $V_{\theta} [\hat{\Pi}]$. 
Suppose a CR type bound $C^{(0)}_{\theta_I,\theta_N}[W_n]$ for the weighted trace of the MSE matrix is 
given and consider the following inequality:
\be \label{qcrbound0}
\Tr{W_n V_{\theta}[\hat{\Pi}]} \ge C^{(0)}_{\theta_I,\theta_N}[W_n], 
\ee 
where we drop the model dependence $\cM_n$ in the bounds to simplify notation. 
%For example, we take $C^{(0)}_{\theta_I,\theta_N}[W_n]=C_{\theta}[W_n,\cM_n]$.
We split the MSE and the weight matrix according to parameter partition $\theta=(\theta_I,\theta_N)$ as before:
\be\nonumber
V_\theta[\hat{\Pi}]=\left(\begin{array}{cc}V_{\theta_I} & V_{\theta_I\theta_N} \\[0.1ex] 
V_{\theta_N\theta_I}& V_{\theta_N}\end{array}\right),\quad
W_n=\left(\begin{array}{cc}W_{I} & W_{IN} \\[0.1ex] 
W_{NI}& W_{N}\end{array}\right).  
\ee
Here the off-diagonal block matrices satisfy $W_{IN}^{\mathrm T}=W_{NI}$ 
and $V_{\theta_I\theta_N}^{\mathrm T}=V_{\theta_N\theta_I}$. 
Rewrite bound \eqref{qcrbound0} as
\begin{multline}\nonumber
\Tr{W_{I}V_{\theta_I}}\ge C^{(0)}_{\theta_I,\theta_N}[W_n]\\
-\Tr{W_{N}V_{\theta_N}}-2\Tr{W_{NI}V_{\theta_I\theta_N}}. 
\end{multline}
Since this holds for all weight matrices $W_n$, which is positive-definite, 
we can evaluate the maximum of the right hand side to define 
\begin{multline} \label{qcrbound1}
C^{\mathrm{Nui}}_{\theta_I}[W_{I},V(\theta_N)]:= 
\max_{\substack{W_{N},W_{IN}:\\ W_n>0}}\big\{ C^{(0)}_{\theta_I,\theta_N}[W_n]\\
-\Tr{W_{N}V_{\theta_N}}-2\Tr{W_{NI}V_{\theta_I\theta_N}} \big\}, 
\end{multline}
where $V(\theta_N)$ represents $V_{\theta_I\theta_N}=(V_{\theta_N\theta_I})^{\mathrm{T}}$ and $V_{\theta_N}$.
Thus, we obtain the following bound for the weighted trace of the MSE about the parameter of interest as
\be
\Tr{W_{I}V_{\theta_I}}\ge C^{\mathrm{Nui}}_{\theta_I}[W_{I},V(\theta_N)]. 
\ee
It is clear that this inequality is achievable if we start with an achievable bound $C^{(0)}_{\theta_I,\theta_N}[W_n]$ 
and the maximum for the optimization \eqref{qcrbound1} exists. 
A natural interpretation of this result is that the MSE matrix must satisfy this inequality 
irrespective of the choice of weight matrix for the nuisance parameter. 

An important remark regarding the positivity condition for the weight matrix. 
As we emphasize in Sec.~\ref{sec:Qnui1}, we impose the condition $W_n>0$ 
in the maximization problem \eqref{qcrbound1}. 
It happens in some case that a maximum does not exists and 
we need to evaluate the supremum. If this is the case, we say that the bound is not explicitly achievable. 
To avoid nonexistence of a maximum, we can then relax the positivity condition as follows. 
\begin{multline} \label{qcrbound1_2}
{\overline{C}}^{\mathrm{Nui}}_{\theta_I}[W_{I},V(\theta_N)]:= 
\max_{\substack{W_{N},W_{IN}:\\ W_n\ge0,\,W_I>0}}\big\{ C^{(0)}_{\theta_I,\theta_N}[W_n]\\
-\Tr{W_{N}V_{\theta_N}}-2\Tr{W_{NI}V_{\theta_I\theta_N}} \big\}. 
\end{multline}
Here the positivity condition applies only for the weight matrix for the parameters of interest. 

We note that the maximization to get the bound $C^{\mathrm{Nui}}_{\theta_I}[W_{I},V(\theta_N)]$ 
seems hard in general and we propose the following bound. 
Instead of maximizing $W_{IN}$ and $W_N$ under the matrix condition $W_n>0$, 
we set the weight matrix as the block diagonal matrix by imposing $W_{IN}=0$. 
This is to look for the maximum within a restricted class of weight matrices. 
Therefore, we can bound the maximization from below as  
\begin{align} \label{qcrbound2}
&C^{\mathrm{Nui}}_{\theta_I}[W_{I},V(\theta_N)]\ge \tilde{C}^{\mathrm{Nui}}_{\theta_I}[W_{I},V_{\theta_N}]\\\nonumber
&:=\max_{W_{N}>0}\big\{ C^{(0)}_{\theta_I,\theta_N}[W_n=W_{I}\oplus W_{N}]-\Tr{W_{N}V_{\theta_N}}\big\}. 
\end{align}
Summarizing the above result, bound \eqref{qcrbound1} (or the approximated one \eqref{qcrbound2}) 
represents what we can do best upon estimating the parameter of interest $\theta_I$ 
in the presence of nuisance parameter $\theta_N$. 
The proposed method of eliminating the nuisance parameters is thus 
based on the similar philosophy to derive the bound \eqref{ccrineq2} in the classical case. 
It is straightforward to observe that the above optimization \eqref{qcrbound1} 
is equivalently expressed as the alternative one: 
\begin{equation} \label{qcrbound3}
\min_{W_{N},W_{IN}:\, W_n>0}\left\{ \Tr{W_nV_\theta[\hat{\Pi}]}-C^{(0)}_{\theta_I,\theta_N}[W_n] \right\}.
\end{equation}

Below we list several remarks concerning the above bound. 
First, the proposed bound \eqref{qcrbound1} depends on 
the estimation errors concerning the nuisance parameter $V_{\theta_I\theta_N}$ and $V_{\theta_N}$. 
This is in contrast to the classical case where the bound in the presence of nuisance parameter is 
solely expressed in terms of the given model, i.e., the partial Fisher information. 
One reason for this dependence is that nuisance parameters enter in POVMs as noted before. 
There is a tradeoff relation for the estimation errors between $\theta_I$ and $\theta_N$ in general. 
Therefore, bound \eqref{qcrbound1} includes this kind of tradeoff (see Sec.~\ref{sec:Qnui4}). 
Alternative interpretation is to regard the dependence of  $V_{\theta_I\theta_N}$ and $V_{\theta_N}$ 
as a biased effect. It is well known in classical statistics that the CR bound 
for non-unbiased estimators does depend on estimation errors in general. 
%(see the generalized CR inequality \eqref{crgen} in Appendix \ref{sec:AppCstat1}) 

Second, it is important to discuss (asymptotic) achievability of the derived bound.  
This is needed especially when the bound $ \tilde{C}^{\mathrm{Nui}}_{\theta_I}[W_{I},V_{\theta_N}]$ 
is used since this bound is obtained within a restricted class of weight matrices. 

%Third, as remarked in the classical case, the bound $C^{\mathrm{Nui}}_{\theta_I}[W_{I},V(\theta_N)]$ 
%depends on the value of unknown nuisance parameter $\theta_N$ in general. 
%So we may get rid of its dependence for example by the worst case estimate: 
%\[
%\max_{\theta_N\in\Theta_N}\ C^{\mathrm{Nui}}_{\theta_I}[W_I,V_{\theta_N}],
%\]
%or by the marginalization (Bayes): 
%\[
%\int_{\Theta_N}\! d\theta_N\ \pi(\theta_N) C^{\mathrm{Nui}}_{\theta_I}[W_I,V_{\theta_N}],
%\]
%where $\pi(\theta_N)$ is a given prior distribution for the nuisance parameter. 

Third, if we start with a bound that is not achievable, the obtained bound according to 
the above procedure is also not achievable in general. For example, let us 
consider an $n$-parameter model with the parameter partition $\theta=(\theta_I,\theta_N)$ as before. 
The symmetric logarithmic derivative (SLD) CR inequality and bound are defined by
\begin{align}
\Tr{W_n V_{\theta}[\hat{\Pi}]}&\ge C^S_\theta[W_n],\\ \nonumber
C^S_\theta[W_n]&=\Tr{W_n G_\theta^{-1}}, 
\end{align}
respectively, where $G_\theta$ is the SLD Fisher information matrix, which is defined by Eq.~\eqref{sldrld} 
in Appendix \ref{sec:AppQFI}. 
It is known that this bound is not achievable unless all SLD operators commute with each other. 
Let us evaluate the approximated bound \eqref{qcrbound2} to see that 
our proposal reduces to a simple result. By splitting the 
inverse of the SLD Fisher information into the block matrix 
analogously to the classical case \eqref{fisherblock}, we get 
\begin{align}\nonumber
&\tilde{C}^{\mathrm{Nui}}_{\theta_I}[W_{I},V_{\theta_N}]\\ \nonumber
&=
\sup_{W_{N}>0}\big\{\Tr{W_I G^{\theta_I\theta_I}}+ \Tr{W_{N}(G^{\theta_N\theta_N}-V_{\theta_N})}\big\}\\
&=\Tr{W_I G^{\theta_I\theta_I}}. 
\end{align}
Here we use the fact $\inf_{W>0} \Tr{WA}=0$ 
for a given positive semi-definite matrix $A$ and $V_{\theta_N}-G^{\theta_N\theta_N}\ge0$. 
Thus, we obtain the same expression as the classical case by replacing the Fisher information 
matrix by the SLD Fisher information matrix. 
In fact, we can work out the detailed analysis without involving the block diagonalization assumption of the weight matrix, 
and then we can show that the minimum exists (see Appendix \ref{sec:AppSuppC}). 

Last, as discussed in the classical case, we can define the information loss 
due to the presence of nuisance parameters for the quantum case. 
Consider an $m$-parameter model $\cM_m$ where all nuisance parameters are known. 
Assume that we have a bound $C_{\theta_I}[W_I,\cM_m]$ for this model, then the difference
\be\label{qinfoloss}
\Delta C_{\theta_I}[W_I]=C^{\mathrm{Nui}}_{\theta_I}[W_{I},V(\theta_N)]-C_{\theta_I}[W_I,\cM_m],
\ee
measures how much information we lose by not knowing the nuisance parameter. 
Unlike the classical case, it is not obvious to derive the condition of $\Delta C_{\theta_I}[W_I,\theta_N]=0$ 
in terms of a given model and weight matrix $W_I$. 
Another difference is that the orthogonal condition does not provide a direct consequence 
for the zero loss of information. One obvious reason is that the concept of 
orthogonality depends on the Fisher information and hence it is not unique 
in the quantum case. The other is that a precision bound is not in general expressed 
as a simple closed-form in terms of quantum Fisher information. 
These will be discussed in the subsequent sections. 

%%%%%%%%%%%%%
\subsection{Achievability of bounds and tradeoff relation}\label{sec:Qnui4}
In this subsection, we discuss achievability of CR tyep bounds and tradeoff relations 
for MSEs between the parameter of interest and the nuisance parameter. 
\subsubsection{Achievability of bounds}
For the parameter estimation problems, we need to be clear in what sense 
a bound is achievable. Consider an $n$-parameter problem and let us minimize 
the weighted trace of the MSE matrix under the locally unbiasedness condition. 
If the optimization problem: 
\[
\min_{\hat{\Pi}\mathrm{\,:l.u.at\,}\theta}\Tr{W_nV_\theta[\hat{\Pi}]}, 
\]
admits the minimizer $\hat{\Pi}_*$, we say that 
the CR type bound $C_\theta[W_n,\cM_n]$ defined by Eq.~\eqref{qcrbound} 
is achievable. If only an infimum for the above minimization exists, on the other hand, 
we say that the bound is achievable in infimum to distinguish two cases.  
The latter happens, for example, when we consider a low-rank weight matrix. %, 
%see Theorem \ref{thmAN} in Sec.~\ref{sec:1para1}. 
There no minimum exists, but only the infimum exists. 

Next, we further distinguish achievable bounds into two cases. 
When an optimal POVM $\Pi_*$ of the optimizer $\hat{\Pi}_*$ depends on the unknown parameter, 
it is impossible to realize this POVM. In this case, we need to utilize 
other methods to attain this bound adaptively in the infinite sample size limit. 
In this paper, we say that a bound is achievable adaptively. 
Note that this paper only deals with bounds for the finite sample size case 
and this is not confused with the asymptotically achievable bound in the usual statistics. 
If an optimal POVM is independent of parameters, we say that the bound is explicitly achievable. 

In the presence of nuisance parameters, achievability problem is much more complicated. 
In particular, we should be carefully distinguish four cases where 
an optimal POVM depends on $\theta_I$ and/or $\theta_N$ or not. 
When $\Pi_*$ depends on $\theta_N$, we must acquire some information 
about the nuisance parameter even though it is of no interest. 

Theorem \ref{thm_equiv} proves that the most natural bound \eqref{qcrboundnui0} 
for locally unbiased estimators about the parameters of interest 
results in equivalent expressions \eqref{qcrboundnui1} or \eqref{qcrboundnui2}. 
From our discussion, we observe that this bound \eqref{qcrboundnui0} is 
tight in the sense of achievability discussed above 
when an optimal estimator for this bound exists. 
%And further, it is also locally unbiased for all parameters or is independent of 
%the nuisance parameters. Otherwise, we should clarify construction of 
%such estimators to realize it in the infinite sample size limit. 

\subsubsection{Tradeoff relation}
Related to the issue mentioned in achievability of bounds, 
an intrinsic nature of the nuisance parameter problem in quantum systems is 
a tradeoff relation between estimating $\theta_I$ and $\theta_N$. 
It is clear that we do not know the value of $\theta_N$, 
and we only want to minimize the MSE about the parameter of interest $\theta_I$. 
However, an optimal measurement to extract information about $\theta_I$ 
requires exact knowledge of $\theta_N$ in general. 
Therefore, we face a problem of designing experiments in such a way that 
we need to extract both $\theta_I$ and $\theta_N$. 
This tradeoff relation is naturally included in our formalism developed in Sec.~\ref{sec:Qnui3} 
and the bound \eqref{qcrbound1}. 

\section{One-parameter model with nuisance parameters}\label{sec:1para}
In this section we focus on models with a single parameter of interest in presence of nuisance parameter(s). 
This class of models is important when applying our method to one-parameter quantum metrology in the presence of quantum noises. 
%which are not known completely. 
It happens that this case is rather special, since the CR type bound and the optimal estimator 
have been known in literature for some time, see for example, Ch.~7 of Ref.~\cite{ANbook}. 

\subsection{One-parameter model with nuisance parameters}\label{sec:1para1}
Consider an $n$-parameter model with $n-1$ nuisance parameters, 
i.e., $\theta_I=\theta_1$ and $\theta_N=(\theta_2,\theta_3,\dots,\theta_n)$.  
Denote this model as 
\be\nonumber
\cM_n=\{ \rho_{\theta=(\theta_1,\theta_N)}\,|\,\theta\in\Theta\}. 
\ee
%We note that this model $\cM_n$ is reduced to the single parameter model $\cM_1$ 
%if all nuisance parameters are completely known. 
%We stress that there are no general formulas for achievable bounds for this class of general models. 
A key result is now given by the following Corollary, which is an extension of 
% for the one-parameter estimation problem in the presence of nuisance parameters(s). 
%The following fundamental theorem also establishes the optimality of the SLD quantum Fisher information matrix, 
%that is its operational meaning 
the theorem [Eq.~(7.93)] in Ch.~7 of Ref.~\cite{ANbook}. 
\begin{corollary}\label{thmAN2}
Given an $n$-parameter regular model $\cM$, 
%suppose that we are interested in estimating the parameter $\theta_1$ 
%in the presence of the nuisance parameters $\theta_N=(\theta_2,\dots,\theta_n)$. 
the achievable lower bound for the MSE about the parameter of interest $V_{\theta_I}[\hat{\Pi}_I]$ is given by 
\be \label{vCRbound2}
\min_{\hat{\Pi}_I\in\cE_{\theta_I}} V_{\theta_I}[\hat{\Pi}_I] = (G_\theta^{-1})_{11}=G^{\theta_I\theta_I}, 
\ee
where the minimization is taken over all locally unbiased estimators $\hat{\Pi}_I$ for the parameter of interest at $\theta$. 
An optimal measurement is given by the projection measurement about the operator:
\be\label{1paraoptPOVM}
L_{\theta}^1=\sum_{j=1}^n g_{\theta}^{1j} L_{\theta,j}. 
\ee
\end{corollary}
%
%We remark that this is a stronger variant of Theorem \ref{thmAN}. 
%In the previous discussion, it was proven only for the infimum 
%of the MSE about the parameter of interest $V_{\theta_I}[\hat{\Pi}]$ under 
%the condition of locally unbiased estimators for all parameters $\theta=(\theta_I,\theta_N)$. 
%The proof for this theorem is given in Appendix \ref{sec:AppPr4}. 

\subsection{Special case}
We analyze the optimal POVM in Corollary \ref{thmAN2} %\ref{thmAN} and \ref{thmAN2} 
and compare it with the optimal one for the case of without the nuisance parameters. 
For two SLD operators, $L_{\theta,1}$ and $L_{\theta}^1:=\sum_{j=1}^n g_{\theta}^{j1} L_{\theta,j}$, 
consider the spectral decomposition by
\begin{align*}
L_{\theta,1}&=\sum_{x\in\cX_1}\lambda_{\theta,1}(x) E_{\theta,1}(x),\\
L_{\theta}^1&=\sum_{x\in\cX^1}\lambda_\theta^{1}(x) E_{\theta}^1(x). 
\end{align*} 
Define the following projections measurements: 
\begin{align} 
\Pi_* &=\{ E_{\theta,1}(x)\}_{x\in\cX_1},\\
\Pi_{\theta_I}^*&=\{ E_{\theta}^1(x)\}_{x\in\cX^1}. 
\end{align}
Since all the SLD operators $L_{\theta,i}$ are linearly independent under our assumptions, 
two measurements $\Pi_*$ and $\Pi_{\theta_I}^*$ become identical 
if and only if the SLD Fisher information matrix is block diagonal 
with respect to the partition $\theta=(\theta_1,\theta_N)$.  
When this orthogonality condition is satisfied, $g_{\theta,11}^{-1}=g_\theta^{11}$ holds. 

First, when all the nuisance parameters are known, we can perform 
the optimal PVM $\Pi_*$ whose Fisher information satisfies 
to $J_\theta[\Pi_*]=g_{\theta,11}$. %by Lemma \ref{lem_optPVM}. 
Therefore, we can attain the SLD CR bound. %\eqref{sldCR_1para}. 
In the presence of nuisance parameters, however, this PVM 
is no longer optimal in general. 
The optimal PVM for estimating the parameter of interest is $\Pi_{\theta_I}^*$ according to Corollary \ref{thmAN2}. 
Since we have information loss \eqref{qinfoloss} due to the nuisance parameter as
\[
\Delta C_{\theta_I}=g_\theta^{11}-(g_{\theta,11})^{-1}\ge0, 
\]
with equality if and only if $\theta_I=\theta_1$ is orthogonal to $\theta_N$, 
the effect of nuisance parameter is not negligible. 
From our discussions, it is clear the following sufficient condition 
suppress effects of the nuisance parameter. 
\begin{align} \nonumber
\mathrm{(i)}&\ \theta_1\mbox{ is globally orthogonal to } \theta_N\\ \label{pocond1}
\mathrm{(ii)}&\ \Pi(L_{\theta_I}^*)\mbox{ is independent of }\theta_N\\ \nonumber
\Leftrightarrow& \forall x\in\cX^1, E_{\theta}^1(x)\mbox{ is independent of }\theta_N, 
\end{align}

For more general case, an optimal measurement depends not only 
on the parameter of interest but the nuisance parameter as well. 
In this case, achievability of the bound is not trivial. 
This will be the subject discussed in the subsequent paper \cite{jsprep}.

%%%%%%%%%%%%%%%%%%%%%%%%%%%%%%%%%%%%%
\section{Qubit models with nuisance parameter}\label{sec:Qbit}
In this section we consider all possible mixed-state models with nuisance parameters 
when the dimension of the Hilbert space is two, that is, qubit models. 
%In the qubit case, the achievable bounds for locally unbiased estimators are 
%known \cite{nagaoka89,hayashi97,GM00}. Then we can apply our proposed bound to derive 
%the MSE bound for the parameter of interest. 
First, note that the possible maximum number of parameters for the qubit models is three. 
Otherwise, the quantum Fisher information matrix becomes singular and is not invertible. 
%Such singular models will not be discussed in this paper. 
%Since the minimum number of parameters in the presence of nuisance parameter is two, 
We thus have the three generic classes of possible qubit models:
\begin{align*}
\mbox{$1+1$ model:}&\ \theta_I=\theta_1,\ \theta_N=\theta_2,\\ 
\mbox{$1+2$ model:}&\ \theta_I=\theta_1,\ \theta_N=(\theta_2,\theta_3),\\
\mbox{$2+1$ model:}&\ \theta_I=(\theta_1,\theta_2),\ \theta_N=\theta_3. 
\end{align*}
We shall call the above models as $1+1$, $1+2$, and $2+1$ qubit models, respectively,  
and these three models will be analyzed in this section. 
Relevant most informative bounds for the qubit case are known, 
and they are given in terms of the SLD Fisher information matrix $G_\theta$. 

\subsection{$1+1$ qubit model}
The achievable CR type bound for two-parameter qubit models 
is the Nagaoka bound \cite{nagaoka89}:
\be\label{nbound}
C_\theta^N[W_2]:=\Tr{W_2G_\theta^{-1}}+2\sqrt{\det W_2G_\theta^{-1}},  
\ee
where $G_\theta$ is the SLD Fisher information matrix. 
This bound holds for all locally unbiased estimators $\hat{\Pi}$ at $\theta=(\theta_1,\theta_2)$. 
We stress that the Nagaoka bound is strictly larger than the SLD CR bound 
since the second term in Eq.~\eqref{nbound} is positive for $W_2>0$. 
Nagaoka also gave an explicit construction of an optimal estimator \cite{nagaoka91}.

We first examine the weight-matrix limit method, which gives bound \eqref{qcrboundnui2}. 
Since $\det W_2$ vanishes, this method yields the following bound for the parameter of interest: 
\be\nonumber
\lim_{W_2\to W_I}  C_\theta^N[W_2]=(G_\theta^{-1})_{11}= g_\theta^{11},  
\ee
where no weight matrix enters in the final expression 
and we use the limit $W_2\to (1,0)^{\mathrm T}(1,0)$. 
This expression is exactly same as the classical case except for 
having the SLD Fisher information instead of the Fisher information. 
From Corollary \ref{thmAN2}, this is also the optimal bound for estimating $\theta_I=\theta_1$. 

Next, we analyze the proposed method by eliminating the nuisance parameter from the Nagaoka bound. 
To perform the maximization, 
we parametrize the weight matrix $W_2$ by 
\be\nonumber
W_2=\left(\begin{array}{cc}w_{11} &w_{12}  \\[0.1ex] 
w_{21}& w_{22}\end{array}\right)=  \left(\begin{array}{cc}w_1 &\sqrt{w_{1}w_{2}}\epsilon  \\[0.1ex] 
\sqrt{w_{1}w_{2}}\epsilon& w_2\end{array}\right),  
\ee
where the positivity conditions are $w_{1},w_{2}>0$ and $|\epsilon|<1$. 
With this form of the weight matrix, we rewrite the maximization as follows. 
We introduce a positive variable $\delta=(w_{2}/w_{1})^{1/2}$, 
and the problem is to maximize $C^{\mathrm{Nui}}_{\theta_I}[W_I]$ over $\delta$ and $\epsilon$. 
Or equivalently, we analyze $C^{\mathrm{Nui}}_{\theta_I}[w_{11}]/w_{11}$ as follows. 
\begin{align} \nonumber 
\frac{1}{w_{11}}&C^{\mathrm{Nui}}_{\theta_I}[W_I]=
\max_{W>0}\frac{1}{w_{11}}\big\{C_\theta^N[W_2]-2 w_{12}V_{\theta_I\theta_N} -w_{22}V_{\theta_N} \big\}\\ \nonumber
&=g_\theta^{11}+
\max_{W_2>0}\big\{-(V_{\theta_N}-g_\theta^{22})\frac{w_{2}}{w_{1}}\\ \nonumber
&-2 \sqrt{\frac{w_{2}}{w_{1}}}\epsilon(V_{\theta_I\theta_N}-g^{12}_\theta) 
+2\sqrt{\det G_\theta^{-1}} \sqrt{\frac{w_{2}}{w_{1}}-\frac{w_{2}}{w_{1}}\epsilon^2} \big\}\\ \nonumber
&=g_\theta^{11}+
\max_{\delta>0,|\epsilon|<1}\big\{-(V_{\theta_N}-g_\theta^{22})\delta^2\\ \nonumber
&-2 \delta\epsilon(V_{\theta_I\theta_N}-g^{12}_\theta) 
+2\sqrt{\det G_\theta^{-1}} \delta\sqrt{1-\epsilon^2} \big\}\\ \nonumber
&=g_\theta^{11}+
\max_{|\epsilon|<1}\frac{\big[ \sqrt{\det G_\theta^{-1}}\sqrt{1-\epsilon^2} 
+(V_{\theta_I\theta_N}-g^{12}_\theta)\epsilon \big]^2}{V_{\theta_N}-g_\theta^{22}}\\  \label{11qnui}
&=g_\theta^{11}+
\frac{\det G_\theta^{-1}
+(V_{\theta_I\theta_N}-g^{12}_\theta)^2}{V_{\theta_N}-g_\theta^{22}}. 
\end{align} 
The first maximization over $\delta>0$ is simply due to a quadratic function of $\delta$ 
and the second one about $\epsilon$ can be done by the Cauchy-Schwarz inequality. 
That is $(a\sqrt{1-\epsilon^2}+b\epsilon)^2\le(a^2+b^2)(1-\epsilon^2+\epsilon^2)=a^2+b^2$. 

We now analyze the obtained bound $C^{\mathrm{Nui}}_{\theta_I}$ in detail. 
First, this is greater than the bound due to the weight-matrix limit method, 
since the second term in Eq.~\eqref{11qnui} is strictly positive. 
Second, this bound represents a tradeoff relation between the estimation errors 
about $\theta_I$ and $\theta_N$. Bounding $(V_{\theta_I\theta_N}-g^{12}_\theta)^2\ge 0$ gives 
\be \label{11qnui2}
V_{\theta_I} \ge g_\theta^{11}+
\frac{\det G_\theta^{-1}}{V_{\theta_N}-g_\theta^{22}}, 
\ee 
which is symmetric between $\theta_1$ and $\theta_2$. 
Third, this result shows that the estimation error regarding the nuisance parameter $\theta_2$ 
has to diverge in order to suppress the second term of Eq.~\eqref{11qnui}. However, 
if this is the case, we completely lose information about the nuisance parameter. 
Since the optimal POVM depends on the nuisance parameter in general, 
we cannot perform this optimal measurement in this case. 
Fourth, the parameter orthogonality condition (with respect to the SLD Fisher information) 
alone does not guarantee no loss of information upon estimating the parameter of interest. 
This condition in this particular example only states that 
the first term in Eq.~\eqref{11qnui} is equal to $g_{\theta,11}^{-1}$, 
but the second term is still finite value in general. 

Last, the MSE inequality by bound \eqref{11qnui} can be expressed after a little algebra as 
\be \label{11qnui3}
\det(V_{\theta_I}-G^{\theta_I\theta_I})\ge\det G^{\theta_I\theta_I}. 
\ee
We note that this relation itself was known before as follows, see for example Ref.~\cite{wsu11}. 
Combining the Gill-Massar inequality \cite{GM00}:
\be
\Tr{J_\theta[\Pi]G_\theta^{-1}}\le 1\quad\mbox{for all }\Pi:\ \mbox{POVM},
\ee 
and the CR inequality $V_\theta[\hat{\Pi}]\ge J_\theta[\Pi]$ for all (locally) unbiased estimators, 
we get 
\be
\Tr{V_\theta[\hat{\Pi}]G_\theta^{-1}}\le 1. 
\ee
This is equivalent to the tradeoff relationship \eqref{11qnui3}. 
However, the Gill-Massar inequality for higher dimensional case are 
not tight in general, and the above derivation does not lead to achievable bounds. 

We finally examine the loss of information due to the nuisance parameter. 
The precision bound when the precise value of nuisance parameter 
is $(g_{\theta,11})^{-1}$, and hence the relevant information loss is 
\be
\Delta C_{\theta_I}= g_\theta^{11}-(g_{\theta,11})^{-1}
+\frac{\det G_\theta^{-1}+(V_{\theta_I\theta_N}-g^{12}_\theta)^2 }{V_{\theta_N}-g_\theta^{22}}. 
\ee
The first two terms indicate the same structure as the classical case, 
whereas the third term is regarded as a genuine quantum quantity. 
We can bound this term by $(\det G_\theta^{-1})/(V_{\theta_N}-g_\theta^{22})$ 
to conclude that the loss of information due to the estimation error tradeoff can be 
suppressed by making large errors in the nuisance parameter. 
However, this in turn means we have less information about it. 

\subsection{$1+2$ qubit model}
When the qubit model contains three unknown parameters, the achievable bound is given by 
the Hayashi-Gill-Massar (HGM) bound \cite{hayashi97,GM00}. This is defined by 
\be \label{hgmbd}
C^{HGM}_{\theta}[W_3]:=\left(F(G_{\theta}^{-1},W_3)\right)^2,
\ee
where $F(A,B):=\Tr{\sqrt{\sqrt{A}B\sqrt{A}}}$ denotes the fidelity 
between two positive semi-definite operators $A,B$. 
%The necessary and sufficient condition for POVMs attaining the HGM bound 
%is precisely derived by Yamagata \cite{yamagata}:  
%$\Pi_{\mathrm{opt}}$ is optimal if and only if its Fisher information matrix satisfies 
%\be \label{optcond}
%J_{\theta}[\Pi_{\mathrm{opt}}]=\frac{ \sqrt{G_{\theta}} \sqrt{F_{\theta}} \sqrt{G_{\theta}}}{\Tr{\sqrt{F_{\theta}}}}, 
%\ee
%where $J_\theta[\Pi]$ denotes the Fisher information matrix about the probability distribution 
%for the measurement outcomes $p_\theta(x)=\tr{\rho_\theta\Pi_x}$ 
%and $F_{\theta}:=(G_\theta^{-1/2}W_3G_\theta^{-1/2})^{1/2}$. 

For this class of model, it is hard to get a closed expression of the HGM bound 
since it involves calculations of the fidelity between $3\times3$ matrices. 
Below we derive an approximated bound by setting the weight matrix as a block matrix, 
i.e., the approximated bound \eqref{qcrbound2} and then making additional approximations. 
To simplify our result, we analyze the case where the SLD Fisher information is 
block diagonalized according to the parameter partition. This can be done 
by performing the parameter orthogonalization. %procedure described in Sec.~\ref{sec3-5}. 
With these assumptions, the approximated bound \eqref{qcrbound2} is obtained by 
maximizing the $2\times2$ weight matrix $W_N$ as 
\begin{align} \nonumber
\tilde{C}^{\mathrm{Nui}}_{\theta_I}[w_{11}]&=
\max_{W_{N}>0}\big\{ C^{HGM}_{\theta}[W_{I}\oplus W_{N}]-\Tr{W_{N}V_{\theta_N}}\big\}\\ \nonumber
&=\max_{W_{N}>0}\Big\{-\Tr{W_{N}V_{\theta_N}}+ \Big(\sqrt{w_{11}g_\theta^{11}}\\ \nonumber
&+\sqrt{\Tr{W_N G^{\theta_N\theta_N}}+2\sqrt{\det W_NG^{\theta_N\theta_N}} }\Big)^2\Big\}\\ \nonumber
&\ge w_{11}g_\theta^{11}+\max_{W_{N}>0}\Big\{
-\Tr{W_{N}(V_{\theta_N}-G^{\theta_N\theta_N} )}\\ \nonumber
&+2\sqrt{\det W_NG^{\theta_N\theta_N}} \\
&+2 \sqrt{w_{11}g_\theta^{11}}
( 4\det W_NG^{\theta_N\theta_N}    )^{1/4}  \Big\},
\end{align}
where the above inequality is due to the fact that  
$\Tr{WA}\ge 2\sqrt{\det WA}$ holds for all positive matrices $W$ and $A$. 
Since $V_{\theta_N}-G^{\theta_N\theta_N}>0$ holds, we can 
reparametrize the weight matrix $W_N$ as 
$(V_{\theta_N}-G^{\theta_N\theta_N})^{-1/2} W (V_{\theta_N}-G^{\theta_N\theta_N})^{-1/2} $. 
Define the quantity
\be\label{delta12}
\delta_{N}=\sqrt{\frac{\det G^{\theta_N\theta_N}}{\det  (V_{\theta_N}-G^{\theta_N\theta_N})}}, 
\ee
and note $\delta_{N}<1$, we get the following chain of inequalities:  
\begin{align} \nonumber
\tilde{C}^{\mathrm{Nui}}_{\theta_I}[w_{11}]
&\ge w_{11}g_\theta^{11}+\max_{W>0}\Big\{
-\Tr{W}+2\delta_{N}\sqrt{\det W}\\ \nonumber
 &\hspace{2cm}+4 \sqrt{w_{11}g_\theta^{11} \delta_N} (\det W)^{1/4} \Big\}\\  \nonumber
& \ge w_{11}g_\theta^{11}+ \max_{T=\Tr{W}>0}\Big\{
-(1-\delta_{N})T \\\nonumber
&\hspace{2cm}+ \sqrt{8w_{11}g_\theta^{11} \delta_N} \sqrt{T} \Big\}\\  \nonumber
&= w_{11}g_\theta^{11}+\frac{2 w_{11}g_\theta^{11} \delta_{N}}{1-\delta_{N}}\\
&=:w_{11} \hat{C}^{\mathrm{Nui}}_{\theta_I}[V(\theta_N)]
\end{align}
where the second inequality follows from $\Tr{W}\ge 2\sqrt{\det W}$ and 
the monotonicity of a function of the form $f(x)=ax+b \sqrt{x}$ with $a,b>0$. 
Third inequality is due to that of a quadratic function and it is attained by 
$\Tr{W}=(2w_{11}g_\theta^{11} \delta_N)^{1/4}/(1-\delta_{N})^{1/2}$. 
Combining all yields the approximated bound for the parameter of interest as
\begin{align}\nonumber 
V_{\theta_I}[V(\theta_N)]&\ge\frac{1}{w_{11}}\tilde{C}^{\mathrm{Nui}}_{\theta_I}[w_{11},V(\theta_N)]\\ \nonumber
&\ge \hat{C}^{\mathrm{Nui}}_{\theta_I}[V(\theta_N)]\\ \label{12qnui}
&=g_\theta^{11}(1+\frac{2 \delta_{N}}{1-\delta_{N}}). 
\end{align}

The obtained bound \eqref{12qnui} clearly shows that 
the estimation error is greater than the value $g_\theta^{11}$ in the presence of nuisance parameters. 
Since $1>\delta_N>0$, $\hat{C}^{\mathrm{Nui}}_{\theta_I,\theta_N}>g_\theta^{11}$ holds. 
We note in passing that the inequality $\delta_N<1$ has the same structure as 
the tradeoff relation \eqref{11qnui3} for the $1+1$ qubit model. 

Let us evaluate the loss of information due to the nuisance parameter. 
The precision bound $(g_{\theta,11})^{-1}$ when the precise value of nuisance parameter 
is known, and hence the information loss is 
\begin{align}\nonumber
\Delta C_{\theta_I}&= \tilde{C}^{\mathrm{Nui}}_{\theta_I}[w_{11},V(\theta_N)]-(g_{\theta,11})^{-1}\\
&=g_\theta^{11}-(g_{\theta,11})^{-1}+g_\theta^{11}\frac{2 \delta_{N}}{1-\delta_{N}} . 
\end{align}
The first two terms exhibit the same structure as the classical one, 
and the last term corresponds to a quantum effect. 
%Recalling expression \eqref{11qnui2}, this information loss does not 
%take into account any tradeoff relations between off diagonal components in the MSE matrix. 

\subsection{$2+1$ qubit model}
The achievable bound for this class of models is the HGM bound \eqref{hgmbd}. 
The same remark applies here as the case of $1+2$ model. 
For this model, we assume that orthogonality condition holds between $\theta_I=(\theta_1,\theta_2)$ and $\theta_N=\theta_3$. 
We set the weight matrix as a block diagonal one. 
A similar procedure can be carried out to derive the approximated bound as follows. 
\begin{align} \nonumber
\tilde{C}^{\mathrm{Nui}}_{\theta_I}[W_{I}]&=
\max_{w_{33}>0}\big\{ C^{HGM}_{\theta}[W_{I}\oplus w_{33}]-\Tr{w_{33}V_{\theta,33}}\big\}\\ \nonumber
&=\max_{w_{33}>0}\big\{ (\sqrt{C^{N}_{\theta_I}[W_{I}]}+\sqrt{w_{33}g_\theta^{33}} )^2\\ \nonumber
&\hspace{3cm}-\Tr{w_{33}V_{\theta,33}}\big\}\\ \nonumber 
&=C^{N}_{\theta_I}[W_{I}]+\max_{w_{33}>0}\big\{ -\Tr{w_{33}(V_{\theta,33}-g_\theta^{33})}\\\nonumber 
&\hspace{2cm}+2\sqrt{C^{N}_{\theta}[W_{I}]w_{33}g_\theta^{33}} \big\}\\ \nonumber
&=C^{N}_{\theta_I}[W_{I}]\big(1+\frac{g_\theta^{33}}{V_{\theta,33}-g_\theta^{33}}  \big),
\end{align}
where the maximization in the third line follows from that about a quadratic function.  
In the above expression, $C^{N}_{\theta_I}[W_{I}]=\Tr{W_I G^{\theta_I\theta_I}}+2\sqrt{\det W_I G^{\theta_I\theta_I}}$ 
denotes the form of the Nagaoka bound for the parameter of interest.  

The above result indicates a similar structure as the previous two cases by noting 
$\det(V_{\theta_N}- G^{\theta_N\theta_N})=V_{\theta,33}-g_\theta^{33}$ and 
$\det G^{\theta_N\theta_N}=g_\theta^{33}$. 
To suppress the effect of the nuisance parameter, 
we need to make $V_{\theta,33}-g_\theta^{33}$ as large as possible, but this in turn implies 
large estimation errors for $\theta_N$. Further, we observe that the quantity 
$g_\theta^{33}/(V_{\theta,33}-g_\theta^{33})$ ($<1$) takes again the 
form $\det G^{\theta_N\theta_N}/\det (V_{\theta_N}-G^{\theta_N\theta_N})$. 
(c.f. the tradeoff relation \eqref{11qnui3} for the $1+1$ qubit model) 

The most informative bound without any nuisance parameter is the Nagaoka bound: 
$C_{\theta_I}^N[W_I,\cM_2]=\Tr{W_I G_{\theta_I\theta_I}^{-1}}+2\sqrt{\det W_I G_{\theta_I\theta_I}^{-1}}
=\Tr{W_I G^{\theta_I\theta_I}}+2\sqrt{\det W_I G^{\theta_I\theta_I}}$ under the assumption of 
parameter orthogonalization between $\theta_I$ and $\theta_N$. 
Therefore, the information loss in this case is
\begin{align}\nonumber
\Delta C_{\theta_I}&= \tilde{C}^{\mathrm{Nui}}_{\theta_I}[W_I,V(\theta_N)]-C_{\theta_I}^N[W_I,\cM_2]\\
&= C^{N}_{\theta_I}[W_{I}]\frac{g_\theta^{33}}{V_{\theta,33}-g_\theta^{33}}. 
\end{align}

%%%%%%%%%%%%%%%%%%%%%%%%%%%%%%%
\section{Examples}\label{sec:Ex}
In this section, we consider several examples to illustrate our findings. 
We are interested in analyzing the bounds for the MSE matrix.  
The first is the one for locally unbiased estimators about the parameter of interest 
\eqref{qcrboundnui0}, which is equivalent to other bounds \eqref{qcrboundnui1} and \eqref{qcrboundnui2}. 
The other is given by expression \eqref{qcrbound1} proposed in Sec.~\ref{sec:Qnui3}. 
For our convenience, we list them here: 
\begin{align*}
C_{\theta_I}[W_I]&=\min_{\hat{\Pi}_I\mathrm{:\,l.u.\,for\,}\theta_I}\Tr{W_IV_{\theta_I}[\hat{\Pi}_I]},\\
C^{\mathrm{Nui}}_{\theta_I}[W_I]&=\Tr{W_IV_{\theta_I}[\hat{\Pi}]}-\\
&\quad\min_{W_{N},W_{IN}:\, W_n>0}\left\{ \Tr{W_nV_\theta[\hat{\Pi}]}-C^{(0)}_{\theta_I,\theta_N}[W_n] \right\}.
\end{align*}
Here and in the following discussion, we drop the model dependence, since it is clear from the context. 

In the first four models, we analyze a qubit state parametrized 
by the standard Stokes parameters, based on which of them are 
the parameter of interest. In the fourth example, 
we give a simple example in which 
the optimal measurement does depend on the value of nuisance parameter. 
In the last two examples, we analyze a model in an open quantum system, 
which is described by the quantum master equation. 

In these examples, we use the standard Pauli matrices $\sigma_i$ ($i=1,2,3$). 
The method of parameter orthogonalization due to Cox and Reid \cite{cr87} 
is extended to the quantum case to demonstrate it is a useful method. 

\subsection{Two-parameter qubit model}
\be \nonumber
\cM_A=\{ \rho_\theta=\frac{1}{2}(I+\theta_1\sigma_1+\theta_2\sigma_2)\,|\,\theta\in\Theta \} ,  
\ee
where $\theta_1$ is the parameter of interest and $\theta_2$ is the nuisance parameter.
The parameter region $\Theta$ is any subset of $\cB_2=\{ (\theta_1,\theta_2)\in\bbr^2\,|\,(\theta_1)^2+(\theta_2)^2<1 \}$. 
The inverse of SLD Fisher information matrix for this model is
\be \nonumber
G_\theta^{-1}=
\left(\begin{array}{cc}1-\theta_1^2 & -\theta_1\theta_2 \\ -\theta_1\theta_2 & 1-\theta_2^2\end{array}\right), 
%I_2-(\theta_1,\theta_2)^{\mathrm T}(\theta_1,\theta_2), 
\ee
and hence $\theta_1$ and $\theta_2$ are not orthogonal in this model. 

We apply the parameter orthogonalization procedure %described before 
and obtain the following new parametrization: 
\be\nonumber
\theta_1=\xi_1,\ \theta_2=c(\xi_2) \sqrt{1-(\xi_1)^2}, 
\ee
where $c(\xi_2)$ is any differentiable function, which is not constant and $\forall\xi_2,\dot{c}(\xi_2):=dc(\xi_2)/d\xi_2\neq0$. 
With this new parametrization, the corresponding SLD Fisher information is diagonalized as
\be\nonumber
G_\xi^{-1}=
\left(\begin{array}{cc} 1-(\xi_1)^2 & 0  \\ 0  & (\frac{1}{1-c(\xi_2)^2}-(\xi_1)^2)^{-1}\dot{c}(\xi_2)^{-2}\end{array}\right). 
\ee 
The SLD operator for the coordinate $\xi_1$ now becomes
\be\label{sldopt1}
L_{\xi,1}=\frac{1}{1-(\xi_1)^2} (-\xi_1I+\sigma_1),  
\ee
which gives the optimal measurement about $\theta_1$. 
This is the projection measurement about $\sigma_1$, which is 
is independent of the nuisance parameter. 
Since the condition \eqref{pocond1} is now satisfied, the projection measurement along $x$ axis is 
the optimal measurement. The bound is then 
\be
V_{\xi,11}[\hat{\Pi}]\ge 1-(\xi_1)^2=1-(\theta_1)^2, 
\ee
that is to say $g_\theta^{11}=1-(\theta_1)^2$ can be achieved explicitly by a locally unbiased estimator about 
the parameter of interest $\theta_1$. It also happens that this measurement does not 
allow us to construct a locally unbiased estimator for both parameters $\theta=(\theta_1,\theta_2)$. 
This is because the resulting probability distribution does not depend on the nuisance parameter $\theta_2$ at all. 
Hence, the MSE for $\theta_2$ formally diverges with this measurement. 
However, as long as we are only interested in estimating $\theta_1$, this does not matter at all. 
%From the discussion in Sec.~\ref{sec4-1}, %%%%%%%%%%
%we have $C^{\mathrm{Nui}}_{\theta_I}[W_I]> C_{\theta_I}[W_I]=W_I g_\theta^{11}$. 
Since the bound $C_{\theta_I}[W_I]$ is achievable explicitly in this example, 
we conclude that the projection measurement along $x$ axis is the optimal 
among all locally unbiased estimators about the parameter of interest. 

\subsection{Estimating one parameter of the standard Stokes parameters}\label{sec:Ex2}
It is straight forward to extend the previous example to the three parameter case. 
Consider the following family of qubit states, which is parametrized the standard Stokes parameters. 
\be\nonumber%\label{qmodel1}
\cM_B=\{ \rho_\theta=\frac{1}{2}(I+\theta_1\sigma_1+\theta_2\sigma_2+\theta_3\sigma_3 )\,|\,\theta\in\Theta \} .  
\ee
Let us assume that $\theta_1$ is the parameter of interest 
and $\theta_N=(\theta_2,\theta_3)$ are the nuisance parameters. 
The parameter region $\Theta$ is any subset of a three dimensional ball with the radius one: 
$\cB_3=\{ (\theta_1,\theta_2,\theta_3)\in\bbr^2\,|\,(\theta_1)^2+(\theta_2)^2+(\theta_3)^2<1 \}$. 
The inverse of SLD Fisher information matrix for this model is
\be\label{q1sld}
G_\theta^{-1}=
\left(\begin{array}{ccc}1-\theta_1^2 & -\theta_1\theta_2&-\theta_1\theta_3 \\
 -\theta_1\theta_2 & 1-\theta_2^2&-\theta_2\theta_3\\ -\theta_1\theta_3&-\theta_2\theta_3&1-\theta_3^2\end{array}\right). 
%I_3-(\theta_1,\theta_2,\theta_3)^{\mathrm T}(\theta_1,\theta_2,\theta_3). 
\ee
Thus, $\theta_1$ and $\theta_N$ are not orthogonal in this model. 
Following the parameter orthogonalization method, we introduce the new parameterization: 
\be\nonumber
\theta_1=\xi_1,\ \theta_2=c_2(\xi_2) \sqrt{1-(\xi_1)^2},\ \theta_3=c_3(\xi_3) \sqrt{1-(\xi_1)^2},
\ee
where $c_2(\xi_2)$ and $c_3(\xi_3)$ are arbitrary functions satisfying the same condition as in Example of Sec.~\ref{sec:Ex1}. 
The SLD Fisher information matrix in the $\xi$ representation is 
\be\nonumber
G_\xi^{-1}=\left(\begin{array}{ccc}1-(\xi_1)^2 & 0 & 0 \\
0 & \frac{1-c_2^2}{\dot{c}_2^2(1-(\xi_1)^2)} & -\frac{c_2c_3}{\dot{c}_2\dot{c}_3(1-(\xi_1)^2)} \\
0 & -\frac{c_2c_3}{\dot{c}_2\dot{c}_3(1-(\xi_1)^2)}  & \frac{1-c_3^2}{\dot{c}_3^2(1-(\xi_1)^2)}\end{array}\right).
\ee
The relevant SLD operator is exactly same as Eq.~\eqref{sldopt1}. 
Applying the general Corollary \ref{thmAN2}, we calculate 
an optimal measurement from 
\[
L_{\theta_1}=\sum_{j=1}^3g_\theta^{j1}L_{\theta,j}=-\theta_1I+\sigma_1,
\]
to find that the optimal measurement is the $\sigma_1$ measurement. 
We see that this optimal measurement is independent of both the parameter of 
interest and nuisance parameter. Thus, we conclude that 
the SLD CR bound $g_{\theta}^{11}=1-(\theta_1)^2$ can be achieved explicitly 
under the condition of locally unbiasedness about $\theta_I=\theta_1$. 
By analyzing the HMG bound and its limit in the weight matrix, we have 
$C^{\mathrm{Nui}}_{\theta_I}[W_I]>C_{\theta_I}[W_I]=C^{\lim}_{\theta_I}[W_I]=W_I g_{\theta}^{11}$. 
%Summarizing above observations, the following relation holds. 
%The conclusion here is exactly same as in the previous example. 

\subsection{Two-parameter submodel}\label{sec:Ex3}
We next consider a two-parameter model: 
\be\nonumber
\cM_C=\{ \rho_\theta=\frac{1}{2}(I+\theta_1\sigma_1+\theta_2\sigma_2+\theta_0\sigma_3 )\,|\,\theta\in\Theta \} .  
\ee
Here, unknown parameters are $\theta=(\theta_1,\theta_2)$ and $\theta_0\neq0$ is a fixed parameter. 
That is to say, we have complete knowledge about $\theta_3=\theta_0$ 
in the previous example in Sec.~\eqref{sec:Ex2}. 
We are only interested in estimating the value of $\theta_1$, whereas the nuisance parameter $\theta_2$ 
is of no interest. 
What is the best estimation strategy in this case? 
At first thought, one may still performs a projection measurement about $\sigma_1$, 
since this model is a submodel of three-parameter model $\cM_B$ analyzed in the previous subsection. 
However, precise knowledge about $\theta_3=\theta_0$ can lower 
the estimation error for the parameter of interest as we see below. 

The inverse of the SLD Fisher information matrix is calculated as
\be\nonumber
G_\theta^{-1}=\left(\begin{array}{cc}1 & 0 \\ 0 & 1\end{array}\right)- 
\frac{1}{1-\theta_0^2}\left(\begin{array}{cc}\theta_1^2 & \theta_1\theta_2 \\ \theta_1\theta_2 & \theta_2^2\end{array}\right). 
%I_2-\frac{1}{1-(\theta_0)^2}(\theta_1,\theta_2)^{\mathrm T}(\theta_1,\theta_2). 
\ee
$\theta_1$ and $\theta_2$ are not orthogonal with respect to 
the SLD Fisher information matrix in this model. 
With the new parametrization, 
\be\nonumber
\theta_1=\xi_1,\ \theta_2=c_2(\xi_2) \sqrt{1-(\theta_0)^2-(\xi_1)^2},
\ee
we can diagonalize the SLD Fisher information as
\be\nonumber
G_\xi^{-1}=
\left(\begin{array}{cc} 1-\frac{(\xi_1)^2}{1-(\theta_0)^2} & 0  \\
 0  &\frac{1-c^2}{\dot{c}^2(1-(\theta_0)^2-(\xi_1)^2)} \end{array}\right). 
\ee 
The SLD operator in this model is 
\be\nonumber
L_{\xi,1}=\frac{1}{1-(\theta_0)^2-(\xi_1)^2} (-\xi_1I+(1-(\theta_0)^2)\sigma_1+\xi_1\sigma_3),  
\ee
and this gives the optimal measurement to estimate $\xi_1=\theta_1$. 
It depends only on the parameter $\xi_1$ but not on $\xi_2$. 

Thus, the SLD CR bound for the parameter of interest is
\be\nonumber
g_\theta^{11}=1-\frac{(\theta_1)^2}{1-(\theta_0)^2},
\ee
which is lower than the previous example $1-(\theta_1)^2$. 
An optimal measurement is similarly worked out to find 
\be
\Pi_{\mathrm{opt}}=\left\{\frac12[I\pm \frac{1}{\sqrt{1+\phi_0(\theta_1)^2}}(\sigma_1+\phi_0(\theta_1)\sigma_3) ]  \right\}, 
\ee
where $\phi_0(\theta_1)=\theta_0\theta_1/(1+(\theta_0)^2)$ is a function of unknown parameter $\theta_1$. 
Note that this optimal measurement is independent of the nuisance parameter, 
we can achieve the above bound in the infinite sample size limit by using an appropriate adaptive scheme. 
%Similar analysis as in the previous example shows the following relations among the bounds: 
%$C_{\theta_I}[W_I]>C_{\theta_I}[W_I]=W_Ig_\theta^{11}$. 
We can conclude that the locally unbiased estimator for the parameter of interest 
is useful here to achieve the bound $C_{\theta_I}[W_I]$. 

\subsection{Estimating two parameters for a qubit model} \label{sec:Ex4}
We consider the same model $\cM_B$ as the example in Sec.~\ref{sec:Ex2}, 
but with different parameters of interest. 
%\be\nonumber
%\cM_2=\{ \rho_\theta=\frac{1}{2}(I+\theta_1\sigma_1+\theta_2\sigma_2+\theta^3\sigma_3 )\,|\,\theta\in\Theta \} ,  
%\ee
$\theta_I=(\theta_1,\theta_2)$ are the parameters of interest and $\theta_3$ is the nuisance parameter. 
The inverse of the SLD Fisher information matrix is given by Eq.~\eqref{q1sld}. 
Therefore, The relevant block submatrix of the inverse SLD Fisher information is 
\be\nonumber
G^{\theta_I\theta_I}=
\left(\begin{array}{cc}1-\theta_1^2 & -\theta_1\theta_2 \\ -\theta_1\theta_2 & 1-\theta_2^2\end{array}\right).  
%I_2-(\theta_1,\theta_2)^{\mathrm T}(\theta_1,\theta_2). 
\ee
An immediate question here is whether we can achieve the Nagaoka bound 
for the parameter of interest in the presence of nuisance parameter $\theta_3$. 
Below, we show the following result about the CR bound for the parameters of interest. 
\be\nonumber
C_{\theta_I}[W_I]=\Tr{W_IG^{\theta_I\theta_I}}+2\sqrt{\Det{W_IG^{\theta_I\theta_I}}}.
\ee

To make our discussion clear, it is convenient to utilize the method of parameter orthogonalization. 
Introduce the new parametrization $\xi=(\xi_1,\xi_2,\xi_3)$ by 
\be\nonumber
\theta_1=\xi_1,\ \theta_2=\xi_2,\theta_3=c(\xi_3)\sqrt{1-(\xi_1)^2-(\xi_2)^2},
\ee
where $c(\xi_3)$ is arbitrary function satisfying the same conditions as previous examples. 
In this new parametrization, the inverse of SLD Fisher information takes 
\be\nonumber
G_\xi^{-1}=\left(\begin{array}{cc}G^{\xi_I\xi_I}& 0\\
0&\frac{1-c(\xi_3)^2}{\dot{c}(\xi_3)^2(1-(\xi_1)^2-(\xi_2)^2)} \end{array}\right). 
\ee
where a $2\times 2$ matrix $G^{\xi_I\xi_I}$ is 
\be\nonumber
G^{\xi_I\xi_I}=
\left(\begin{array}{cc}1-\xi_1^2 & -\xi_1\xi_2 \\ -\xi_1\xi_2 & 1-\xi_2^2\end{array}\right). 
%I_2-(\xi_1,\xi_2)^{\mathrm T}(\xi_1,\xi_2)=G^{\theta_I\theta_I}. 
\ee
With this parameter orthogonalization method, the weight-matrix limit immediately gives
$C^{\lim}_{\xi_I}[W]=\Tr{WG^{\xi_I\xi_I}}+2\sqrt{\Det{W G^{\xi_I\xi_I}}}$. 

Two relevant SLD operators for the parameter of interest $\xi_I=(\xi_1,\xi_2)$ are 
\begin{align*}
L_{\xi,1}&=\frac{1}{1-|\xi_I|^2}[-\xi_2 I+ \xi_1\xi_2\sigma_1+(1-(\xi_1)^2)\sigma_2],\\
L_{\xi,2}&=\frac{1}{1-|\xi_I|^2}[-\xi_1 I+ (1-(\xi_2)^2)\sigma_1+\xi_1\xi_2\sigma_2],
\end{align*}
with $|\xi_I|^2=(\xi_1)^2+(\xi_2)^2$. 
They depend only on the parameters of interest. 
From the SLD operators about $\xi_I$, we see that an optimal measurement for $\xi_I$ is solely determined by 
certain combinations of $L_{\xi,1}$ and $L_{\xi,2}$, see for example \cite{yamagata}.  
We then conclude that it is independent of the nuisance parameter $\xi_3$. 
Furthermore, the resulting probability distribution from the optimal measurement 
only depends on the parameters of interest, we can construct a locally unbiased 
estimator for the parameter of interest at any point. 
We thus have $C_{\xi_I}[W]=\Tr{WG^{\xi_I\xi_I}}+2\sqrt{\Det{W G^{\xi_I\xi_I}}}$. 
This is, of course, true from the general theorem \ref{thm_equiv}. 
Noting $\theta_I=(\theta_1,\theta_2)=(\xi_1,\xi_2)$, we prove our claim: 
$C_{\theta_I}[W]=C_{\xi_I}[W]$ is the adaptively achievable bound for the parameter of interest.  

\subsection{Nuisance parameter as a qubit phase}\label{sec:Ex5}
In all previous examples, optimal measurements are independent of nuisance parameters. 
And hence, we can safely ignore tradeoff relations between the MSEs 
for the parameter of interest and the nuisance parameter. In this example, 
we provide a simple example in which we cannot ignore the existence of a nuisance parameter. 

Suppose we are interested in estimating the value of $\theta_1$ of 
the following model:
\begin{multline}
\cM_E=\{ \rho_\theta=\frac{1}{2}(I+\theta_1\cos\theta_2\sigma_1+\theta_1\sin\theta_2\sigma_2 )\\
\,|\,\theta_1\in(0,1),\theta_2\in [0,2\pi)\} ,  
\end{multline}
where the phase $\theta_2$ is the nuisance parameter.  
The inverse of the SLD Fisher information matrix of this model is 
\be\nonumber
G_\theta^{-1}=
\left(\begin{array}{cc} 1- (\theta_1)^2& 0  \\[1ex] 0 & (\theta_1)^{-2}\end{array}\right), 
\ee
that is $\theta_1$ and $\theta_2$ is globally orthogonal with respect to the SLD Fisher information. 

When the precise value of the phase $\theta_2$ is known, 
the optimal measurement is given by the projection measurement about the SLD operator:
\be\nonumber
L_{\theta,1}=\frac{1}{1-(\theta_1)^2}(-\theta_1I+\theta_1\cos\theta_2\sigma_1+\theta_1\sin\theta_2\sigma_2 ). 
\ee
That is 
\be
\Pi_{\mathrm{opt}} =\{\frac{1}{2} [I\pm (\cos\theta_2\sigma_1+\sin\theta_2\sigma_2)]  \},
\ee
and the Fisher information matrix about this optimal estimator is 
\be
J_\theta[\Pi]=g_{\theta,11}=[1- (\theta_1)^2]^{-1}. 
\ee

Next, let us consider the case when $\theta_2$ is not known, 
that is $\theta_2$ is a nuisance parameter. 
Since this measurement $\Pi_{\mathrm{opt}}(\theta_2)$ depends on the nuisance parameter $\theta_2$, 
the condition (ii) of the sufficient condition \eqref{pocond1} is not satisfied. 
In other words, there is no way to perform it in the presence of nuisance parameter $\theta_2$. 
We also note that  the measurement $\Pi_{\mathrm{opt}}(\theta_2)$ gives 
a probability distribution with only two outcomes. It is then impossible 
to infer two parameters $\theta_1,\theta_2$. 
The score functions become linearly dependent, and hence, the Fisher information matrix 
is singular. 
In this case we follows the proposed method to get the CR type bound \eqref{11qnui2} 
for any locally unbiased estimator at $\theta$: 
\be
V_{\theta,11}[\hat{\Pi}] \ge g_{\theta}^{11}\Big(1+
\frac{g_{\theta}^{22} +(V_{\theta,12})^2}{V_{\theta,22}-g_{\theta}^{22}}\Big), 
\ee
which is strictly greater than $g_{\theta}^{11}=(g_{\theta,11})^{-1}$. 

This simple model exhibits a unique feature of the nuisance parameter problem in the quantum case. 
Estimating two-parameter qubit state in the presence of unknown phase as a nuisance parameter 
was discussed in Ref.~\cite{js15}, which provides a discussion on asymptotic achievability of the above bound.

\subsection{Open quantum system}\label{sec:Ex6}
In this example, we shall analyze a qubit model given by the solution to the quantum master equation of the form: 
\be
\del_t \rho(t)=\I [\rho(t),H_0]+\frac14\sum_{i=1,2,3}\gamma_i \big[\sigma_i,[\sigma_i,\rho(t)]\big], 
\ee
in the unit of $\hbar=1$. Here, the free Hamiltonian is $ H_0= \omega \sigma_3/2$ 
and we wish to estimate the phase accompanied by the evolution of the system. 
If we specify an initial state as $\v{s}_0=(s_1,s_2,s_3)$ in terms of the Bloch vector 
and for a given later time $t$ is fixed, we can consider a model: 
\be
\cM_F=\{ \rho_{\theta}(t) \,|\,\theta\in\Theta \} ,  
\ee
where $\theta=(\theta_1,\theta_N)$ with $\theta_1=\omega t$ and 
$\theta_N$ represents the damping parameters $\gamma_i$. 

When the precise values of all damping parameters are known, 
the model is a single parameter. The SLD CR bound is 
\be
V_{\theta,11}[\hat{\Pi}]\ge g_{\theta,11}^{-1}.  
\ee
If the precise values are not completely known, 
we can utilize the proposed procedure to derive a CR type bound. 

Let us analyze the simplest case where all damping parameters are equal, 
i.e., an isotropic noise model. Set $\theta_2=\gamma_1 t=\gamma_2 t=\gamma_3 t$
we have a two-parameter model. The SLD Fisher information matrix of this model is 
\be
G_\theta=
\left(\begin{array}{cc} \Exp{-2\theta_2}(s_1^2+s_2^2)& 0  \\ 0  & \ds\frac{\Exp{-2\theta_2}|\v{s}_0|^2}{1-\Exp{-2\theta_2}|\v{s}_0|^2}\end{array}\right),  
\ee
which is diagonal and is independent of the parameter of interest $\theta_1$. 
Here and below, we assume that $s_1^2+s_2^2\neq0$. 

It happens that the projectors about the SLD operator $L_{\theta,1}$ is independent of $\theta_2$ 
and two conditions \eqref{pocond1} are satisfied. 
Therefore, we conclude that the bound 
\be
g_{\theta,11}^{-1}=g_{\theta}^{11}=\Exp{2\theta_2}(s_1^2+s_2^2)^{-1}, 
\ee
can be achieved by performing the following projection measurement: 
\begin{multline}
\Big\{ \frac{1}{2} \big[I\mp\frac{1}{|\v{s}_0|}(s_1\sin\theta_1+s_2\cos\theta_1 )\sigma_1\\
\pm\frac{1}{|\v{s}_0|}(s_1\cos\theta_1-s_2\sin\theta_1 )\sigma_2 \big]\Big\}. 
\end{multline}

It should be emphasized that the above simple result holds 
only for a special class of noise models. One of key ingredients is that 
two processes of unitary and damping 
are completely factorized in the following sense.  
Denoting the state in terms of the Bloch vector $\v{s}(t)$ at later time $t$, 
the dynamics is given 
\be
\v{s}(t)= \Gamma(t) u_\theta(t) \v{s}_0,  
\ee
where $u_\theta(t)$ denotes the rotation according to the free Hamiltonian 
and $\Gamma(t) $ accounts for the damping process, which is a $3\times 3$ matrix. 
If a solution to a given master equation is not factorized or 
the damping matrix $\Gamma(t) $ also depends on the parameter of interest $\theta$, 
we can no longer use the bound $g_{\theta,11}^{-1}$ as the ultimate bound. 
This point will be discussed in the next example. 
%This happens, for example, we have a slight anisotropy in $\gamma_1$ and $\gamma_2$ 
%or the master equation contains non-Markovian effect. This will be discussed in detail 
%in the subsequent paper \cite{jsprep}. 

\subsection{A realistic noise model}%%%%%%%%%%%%%%%%%%
So far, we have provided simple examples in which we cannot ignore the nuisance parameter. 
As the last example, we shall consider a physically motivated model, 
a single-spin model affected by random magnetic fields.  
To simplify our notation we set $\hbar=1$ and the magnetic moment $\mu_B=1$ unless noted otherwise. 

\subsubsection{Model setting}
Consider a frequency estimation problem described by the Hamiltonian $H=H_0+H_{\mathrm{noise}}(t)$. 
$H_0=\omega\sigma_3/2$ generates a unitary evolution and $H_{\mathrm{noise}}(t)$ 
describes the effect of random magnetic fields defined by
\be
H_{\mathrm{noise}}(t)=\frac{1}{2}\sum_{i=1}^3b_i(t)\sigma_i. 
\ee 
Here the time-dependent magnetic field $\v{b}(t)=\big(b_1(t),b_2(t),b_3(t)\big)$ obeys the Maxwell equation 
and is fluctuating with the following time average characteristics: 
\be\nonumber
\overline{b_i(t)}=0,\quad \overline{b_i(t)\,b_j(t')}=\delta_{ij}\,b^2\Exp{-\Gamma|t-t'|} , 
\ee
where the bars denote the time average over a classical probability density function 
(See Ref.~\cite{hnse11} for the derivation and more details). 

We follow the standard quantum master equation approach to derive 
the von Neumann equation for the qubit-state density matrix: 
\begin{align}\label{model}
&\frac{\del}{\del t}\rho(t) =-\I[H_0,\rho(t)] + {\cal L}[\rho(t)],\\ \nonumber
&{\cal L}[\rho(t)]=-\frac\I2 [\delta\omega(t) \sigma_3, \rho(t)]-\frac{1}{4}\sum_{i=1}^3\gamma_i(t)\big[\sigma_i,[\sigma_i,\rho(t)]\big],
%&\delta\omega(t)=b^2\int_0^t dt' \Exp{-\Gamma t'}\sin\omega t',\\\nonumber
%&\gamma_1(t)=\gamma_2(t)=b^2\int_0^t dt' \Exp{-\Gamma t'}\cos\omega t',\\\nonumber
%&\gamma_3(t)=b^2\int_0^t dt' \Exp{-\Gamma t'}. \nonumber
\end{align}
where the time-dependent decaying factors are
\begin{align}\label{model1}
&\delta\omega(t)=b^2\int_0^t dt' \Exp{-\Gamma t'}\sin\omega t',\\\nonumber
&\gamma_1(t)=\gamma_2(t)=b^2\int_0^t dt' \Exp{-\Gamma t'}\cos\omega t',\\\nonumber
&\gamma_3(t)=b^2\int_0^t dt' \Exp{-\Gamma t'}. \nonumber
\end{align}
To get the above equation, we only used the following approximation: 
$\rho(t')\simeq \Exp{-\I H_0(t'-t)}\rho(t)\Exp{\I H_0(t'-t)}$, 
which is valid when the coupling is small $(b/ \omega)^2\ll 1$. 

In the Born-Markov approximation, we further 
set $t\to\infty$ in the upper limits for the integrations to 
get time-independent decaying coefficients: 
\be\label{model2}
%&\frac{\del}{\del t}\rho(t) =-\I[H_0,\rho(t)] + {\cal L}[\rho(t)],\\
%&{\cal L}[\rho(t)]=-\I\frac12 [\Omega(t) \sigma_3, \rho(t)]-\frac{1}{4}\sum_{i=1,2,3}\Gamma_i(t)[\sigma_i,[\sigma_i,\rho(t)]],\\
\delta\omega=\frac{b^2\omega}{\Gamma^2+\omega^2},\ \gamma_1=\gamma_2=\frac{b^2\Gamma}{\Gamma^2+\omega^2},\ \gamma_3=\frac{b^2}{\Gamma}. 
\ee
This approximation is valid when time is much larger than the decoherence time $\Gamma^{-1}$. 

When the noise parameters $(b,\Gamma)$ are much small than $\omega$, we can further simplify the model 
to arrive at a so-called parallel noise model:
\be\label{model3}
%&\frac{\del}{\del t}\rho(t) =-\I[H_0,\rho(t)] + {\cal L}[\rho(t)],\\
%&{\cal L}[\rho(t)]=-\I\frac12 [\Omega(t) \sigma_3, \rho(t)]-\frac{1}{4}\sum_{i=1,2,3}\Gamma_i(t)[\sigma_i,[\sigma_i,\rho(t)]],\\
\delta\omega=0,\ \gamma_1=\gamma_2=0,\ \gamma_3=\frac{b^2}{\Gamma}=:\gamma. 
\ee
This approximation is valid, for example, the strength of the fluctuating magnetic field is of order $b\sim10^{-9}$T 
and the noise correlation is $\Gamma\sim100 $Hz in the usual units such that the relation $b, \Gamma\ll \omega $ holds \cite{hnse11}. 

In the following we compare the above three models described by Eqs. (\ref{model1}, \ref{model2}, \ref{model3}). 
For convenience, we call them model 1, 2, 3, respectively. 
Under the assumption of time $t$ is precisely known, 
Three models are characterized by three parameters $(\theta_1,\theta_2,\theta_3)=(\omega,b^2,\Gamma)$. 
(Model 3 depends essentially on two parameters as $(\omega,\gamma)$ only.) 
Here $\theta_1=\omega$ is the parameter of interest whereas others are nuisance parameters. 
We can explicitly solve the time evolution of the Bloch vector 
for a given initial state $s_0=(s_1,s_2,s_3)$ and it is given by
\begin{align} \label{solution}
&\cM=\{s_\theta(t)=A_{\theta}(t)s_0\,|\,\theta=(\omega,b^2,\Gamma)\in\Theta\},\\ \nonumber
&A_{\theta}(t):=\left(\begin{array}{cc} \Exp{-\Gamma_1(t)}
\ds\binom{\cos\Omega(t) \ \ \sin\Omega(t)}
 {-\sin\Omega(t) \ \cos\Omega(t)}& 0 \\
0 & \Exp{-\Gamma_3(t)}\end{array}\right), \\ \nonumber
%&A_{\theta}(t):=\left(\begin{array}{cc} 
%\Exp{-\Gamma_1(t)}\left(\begin{array}{cc}\cos\Omega(t) & \sin\Omega(t)
% \\-\sin\Omega(t) & \cos\Omega(t)\end{array}\right)&0\\
%  0 & \Exp{-\Gamma_3(t)}\end{array}\right). \\ \nonumber
%&A_{\theta}(t):=\left(\begin{array}{ccc} \Exp{-\Gamma_1(t)}\cos\Omega(t) & \Exp{-\Gamma_1(t)}\sin\Omega(t) &0\\ 
%-\Exp{-\Gamma_1(t)}\sin\Omega (t) & \Exp{-\Gamma_1 (t)}\cos\Omega (t) & 0 \\0 & 0 & \Exp{-\Gamma_3(t)}\end{array}\right). \\ \nonumber
&\Gamma_1(t)=\int_0^tdt'[\gamma_1(t')+\gamma_3(t')],\ \Gamma_3(t)=2\int_0^tdt' \gamma_3(t'),\\ \nonumber
&\Omega(t)=\omega t+\int_0^tdt'\delta\omega(t').
\end{align}
We then compute the SLD Fisher information matrix $G_\theta$. 
The explicit expressions of $G_\theta$ for model 1, 2 are rather complicated and are omitted. 
Model 3 takes a similar form as the previous example in Sec.~\ref{sec:Ex6}.  
%\begin{align} 
%&\cM=\{s_\theta(t)=A_{\omega,\gamma}(t)s_0\,|\,\theta=(\omega,\gamma)\in\Theta\},\\ \nonumber
%&A_{\omega,\gamma}(t):=\left(\begin{array}{ccc} \Exp{-\gamma t}\cos\omega t & \Exp{-\gamma t}\sin\omega t &0\\ 
%-\Exp{-\gamma t}\sin\omega t & \Exp{-\gamma t}\cos\omega t & 0 \\0 & 0 & 1\end{array}\right). 
%\end{align}

In the absence of external noise, the SLD Fisher information about the parameter $\theta=\omega$ is 
$G_\theta =(s_1^2+s_2^2)t^2$. The maximum sensitivity is achieved by 
arbitrary pure initial state on $xy$ plane, and the optimal MSE is $t^{-2}$.  
This observation also holds in model 3 as previously known, 
and the maximum SLD Fisher information matrix with this optimal initial state is 
$(G_\theta)^{-1}_{11}=(G_{\theta,11})^{-1} =\exp(-2\gamma t)  t^{-2} $. Note that 
the effect of nuisance parameter $\gamma$ is irrelevant in model 3 and 
this bound can be achieved asymptotically by the adaptive scheme. 
However, model 2 and 3 do not satisfy the condition of parameter orthogonality with respect to the SLD Fisher information matrix. 
Therefore, further detail analysis is needed to confirm that the general bound 
$(G_\theta)^{-1}_{11}$ is achievable or not. 
%This point shall be discussed in the subsequent publication. 

\subsubsection{Comparison and discussion}
We now compare three models 1, 2, and 3 to see the validity of approximations commonly used in the literature. 
Model 3 is the simplest one and has been widely adopted. 
(C.f., the isotropic noise model in Sec.~\ref{sec:Ex6}, $\gamma_1=\gamma_2=\gamma_3$, is another familiar example of this kind.)  
Model 2 takes into account the effect of random fields up to the second order in the coupling constant $b$. 
This is based on the assumption of small $b/\omega$, and the solution is valid for large $\Gamma t$ regime. 
Model 1 is more general than others, in which the kernel for the master equation $\gamma_i(t)$ is time-dependent. 
Some authors regard this model as non-Markovian in the sense of finite correlation time for the decaying factor. 
In this paper, however, we prefer to call model 3 as the weak-coupling model with a finite memory, 
and this model is valid as long as $b/\omega$ is small. 

It is widely known that a non-Markovian environment may bring back coherence in some cases 
when compared with the Markovian case. In quantum metrology, several authors reported this ``memory effect" 
showing that the SLD Fisher information decays much slower in time than the Markovian case, 
see for example Ref.~\cite{mbf11,chp12,stc14}. 
We now analyze if this memory effect is true even when noise parameters 
are not completely known, that is, they are treated as the nuisance parameters. 

In the following, we compare five different quantities related to the SLD Fisher information 
about the parameter of interest $\theta_1=\omega$. Fix a pure initial state $s_0=(\sqrt{1-z^2},0,z)$ 
and let it evolve according to three noise models. The Bloch vector at later time $t$ for model $q$ ($q=1,2,3$) 
is denoted by $s_\theta^{(q)}(t)$ [Eq.~\eqref{solution}]. We write $(1,1)$ components of the SLD Fisher information matrix 
and the inverse of SLD Fisher information matrix about model $q$ by 
\begin{align}
&g^{(q)}_{11}(\omega,b^2,\Gamma,z,t),\\
&g_{(q)}^{11}(\omega,b^2,\Gamma,z,t),
\end{align}
respectively. The quantity $(g^{(q)}_{11})^{-1}$ represents the precision limit 
when all noise parameters $(b^2,\Gamma)$ are completely known without any uncertainty, i.e., without nuisance parameters. 
The quantity $g_{(q)}^{11}$, on the other hand, is the bound when 
we treat the noise parameters as the nuisance parameters. 
(C.f. its inverse is called the partial SLD Fisher information.) 
The general relationship $g_{(q)}^{11}\ge (g^{(q)}_{11})^{-1}$ holds for all parameters, and 
$g_{(3)}^{11}= (g^{(3)}_{11})^{-1}$ holds as a special case. 

\begin{figure}[H]
\begin{center}
\vspace{-5mm}
\subfigure[]{
\label{fig1a}
\includegraphics[width=0.9\columnwidth]{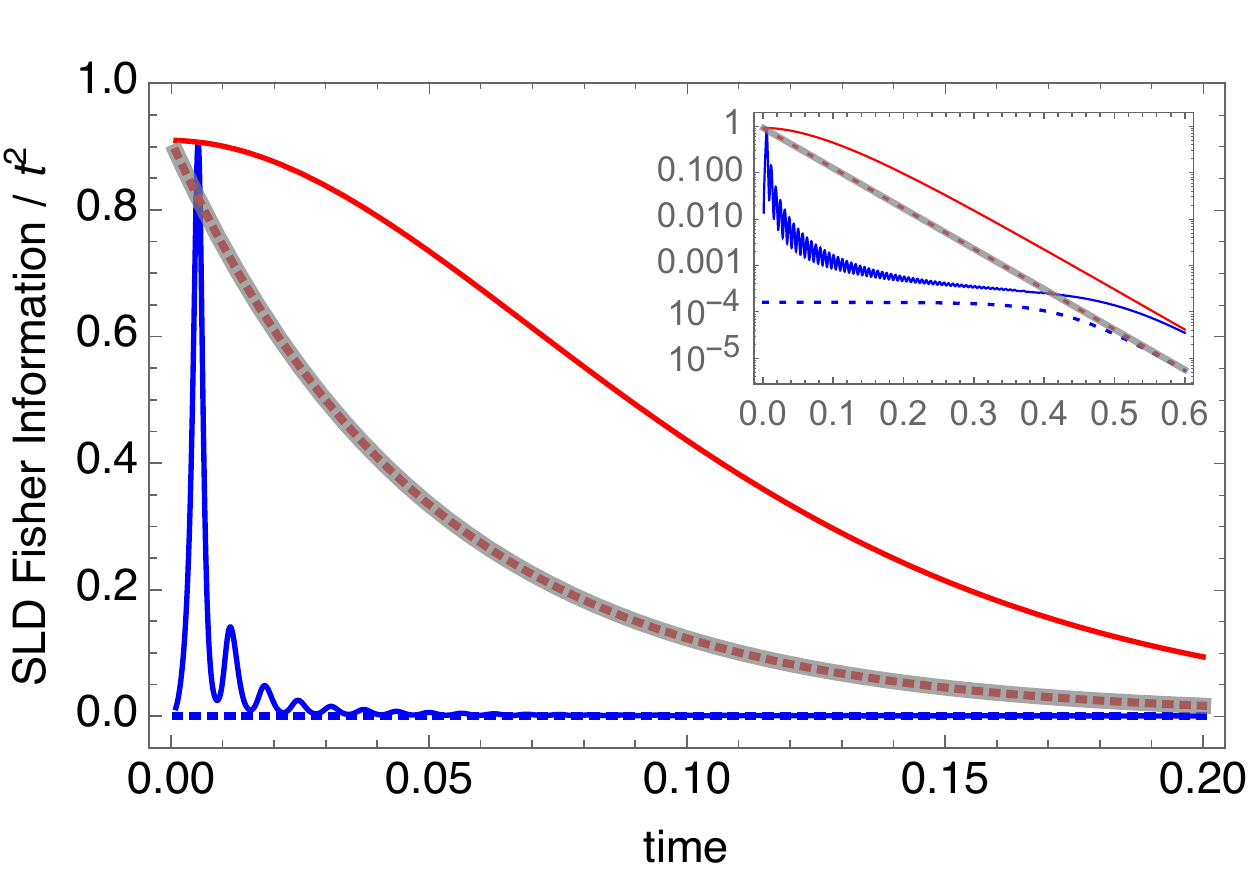}
}
\vspace{-5mm}
\subfigure[]{
\label{fig1b}
\includegraphics[width=0.9\columnwidth]{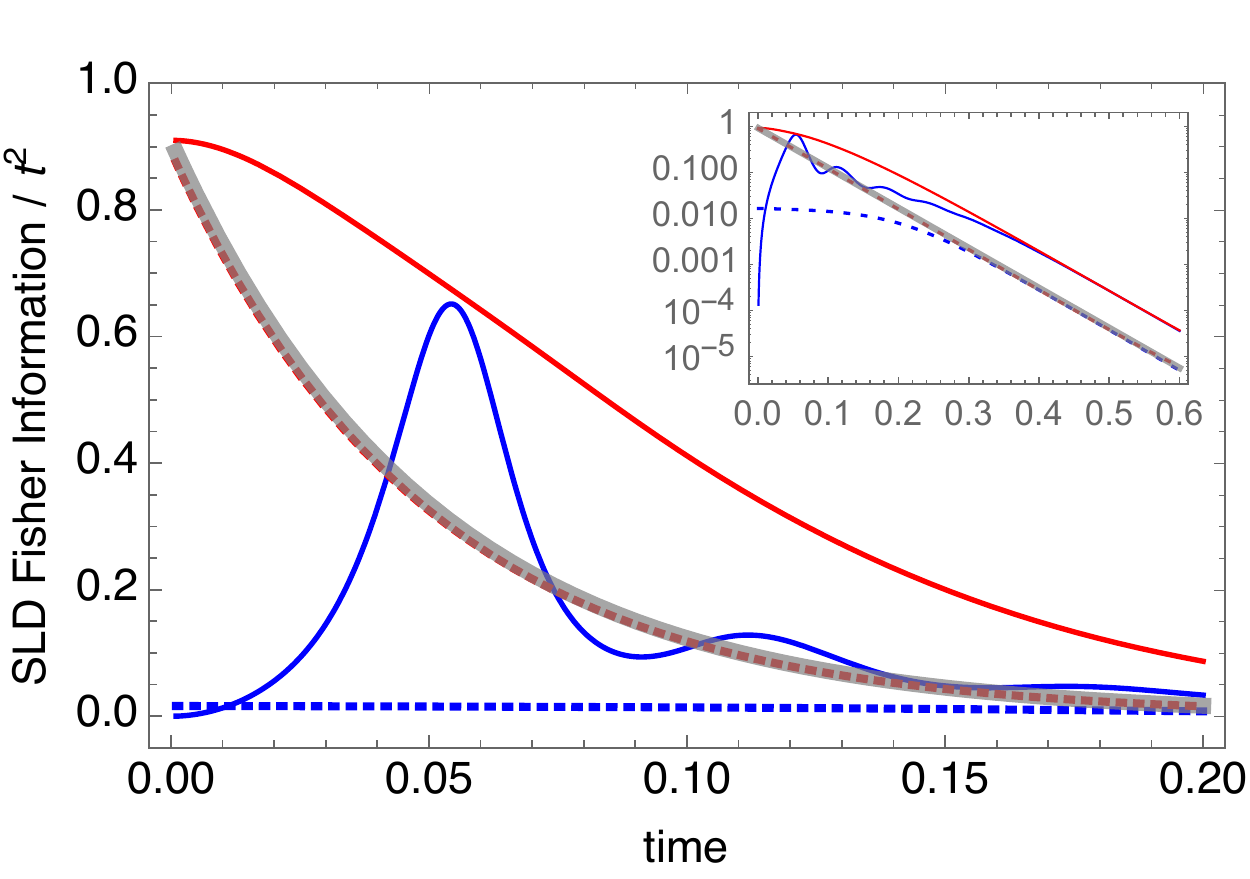}
}
\vspace{-5mm}
\subfigure[]{
\label{fig1c}
\includegraphics[width=0.9\columnwidth]{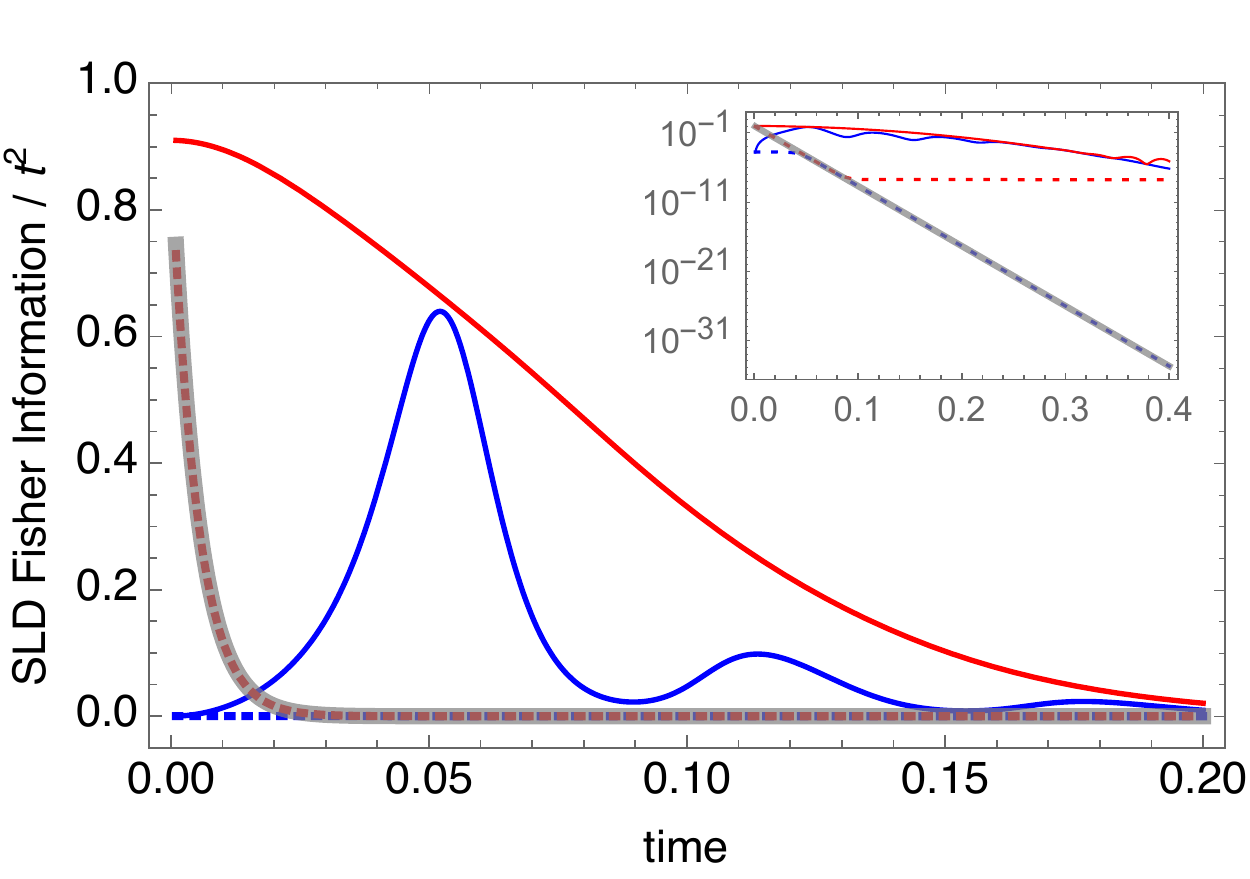}
}
\caption{(Color online) Dynamics of the SLD Fisher information (in units of $t^{-2}$) 
about the parameter of interest $\theta_1=\omega$ for three different approximations (\ref{model1}, \ref{model2}, \ref{model3}). 
Parameters $(\omega,b^2,\Gamma)$ are Fig.~\ref{fig1a}; $(10^3,10^2,10)$, Fig.~\ref{fig1b}; $(10^2,10^2,10)$, 
and Fig.~\ref{fig1c}; $(10^2,10^2,1)$. 
Five curves are model 1; $(g_{(1)}^{11})^{-1}$(Blue solid line), $g^{(1)}_{11}$(Red solid line). 
Model 2; $(g_{(2)}^{11})^{-1}$(Blue dotted line), $g^{(2)}_{11}$(Red dotted line). 
Model 3; $(g_{(3)}^{11})^{-1}=g^{(3)}_{11}$(Gray solid line). Insets show the logarithmic plot.}
\label{fig1}
\end{center}
\end{figure}
In the following, we plot $g^{(q)}_{11}/t^2$ and $(g_{(q)}^{11})^{-1}/t^2$ ($q=1,2,3$) as functions of time 
for fixed parameters $(\omega,b^2,\Gamma)$ and the initial state $z\in[0,1]$. 
When analyzing this problem, we find that the limit $z\to0$ results in 
the zero SLD Fisher information for model 1, 2, and we keep $z$ finite. 
In Fig.~\ref{fig1}, we plot five functions: 
%{$(g_{(1)}^{11})^{-1}$ (solid line)}, {$g^{(1)}_{11}$ (dashed line)}, {$(g_{(2)}^{11})^{-1}$ (dash-dotted line)}, 
%{$g^{(2)}_{11}$ (gray-solid line)}, {$(g_{(3)}^{11})^{-1}=g^{(3)}_{11}$ (dotted line)} 
Model 1; $(g_{(1)}^{11})^{-1}$ (Blue solid line), $g^{(1)}_{11}$ (Red solid line),  
Model 2; $(g_{(2)}^{11})^{-1}$ (Blue dotted line), $g^{(2)}_{11}$ (Red dotted line). 
Model 3; $(g_{(3)}^{11})^{-1}=g^{(3)}_{11}$(Gray solid line). 
Parameter values $(\omega,b^2,\Gamma)$ are set to 
$(10^3,10^2,10)$ for Fig.~\ref{fig1a}, $(10^2,10^2,10)$ for Fig.~\ref{fig1b}, 
and $(10^2,10^2,1)$ for Fig.~\ref{fig1c}, respectively. 
The initial state parameter is chosen as $s_0=(\sqrt{0.91},0,0.3)$. 
The ratio $(b/\omega)^2$ represents the weakness of the coupling, which is to be weak 
in order for the master equation approach to be valid. The values for above two choices 
are $(b/\omega)^2=10^{-4},10^{-2}, 10^{-4}$, that satisfy the weak-coupling condition $(b/\omega)^2\ll1$.  
Insets of the figures show the logarithmic plot for longer time behaviors. 

Let us now discuss the result shown in Fig.~\ref{fig1a}. 
In the absence of nuisance parameters, model 2 and 3 are almost identical 
confirming that model 3 is a good approximation to model 2 within our parameter setting. 
Model 1, which is more accurate than others, has larger SLD Fisher information about 
the parameter of interest $\theta_1=\omega$ for the short time scale. 
This effect is known as the memory effect in literature and is believed to bring 
an advantage upon estimating $\omega$. This observation is seen in our example. 

When the noise parameters $(b^2,\Gamma)$ are treated as the nuisance parameters, on the other hand, 
we have to use $(g_{(q)}^{11})^{-1}$ as the correct SLD Fisher information about the parameter of interest $\theta_1=\omega$. 
In this case, we observe completely different behaviors for model 1 and model 2. 
First, model 2 (blue dotted line) immediately drops down when compared with the case of no nuisance parameters (red dotted line). 
Model 1 (blue solid line), which is valid for the short time scale, exhibits amplification of the SLD Fisher information, 
however, this effect is much weaker than the case of no nuisance parameters (red solid line). 
Figures \ref{fig1b} and \ref{fig1c} also show qualitatively same behaviors as Fig.~\ref{fig1a}. 
The major difference is that the enhancement due to the memory effect is more visible than Fig.~\ref{fig1a}. 
In particular, good amplification is observed in Fig.~\ref{fig1c}. 
The above results show that the effect of nuisance parameters are not negligible even in the weak-coupling limit. 
Furthermore, the memory effect, which is due to a time-dependent kernel for the master equation, 
is an artifact of improper analysis for the problem within our physical model.

%%%%%%%%%%%%%%%%%%%%%%%%%%%%%%%%%%%%%%%%%
\section{Conclusion}\label{sec:Conc}
In this paper, we have formulated the nuisance parameter problem in the quantum estimation theory. 
Unlike the classical case, the estimation error bound for multi-parameter case 
cannot be expressed as a simple formula in terms of the partial quantum Fisher information. 
This is partly due to the intrinsic nature of the problem involving measurement degrees of freedoms. 
We have proposed a bound for estimating the parameter of interest by eliminating the nuisance parameters from a given expression of the bound. 
This is based on the similar philosophy as in the classical case, i.e., the best estimation strategy is to estimate 
all parameters including the nuisance parameter. 
However, we have pointed out that we should not ignore a tradeoff relation between 
the mean square error of the parameters of interest and the nuisance parameters. 
Another key observation in our results is that a class of estimators needs to be 
clarified when dealing with the nuisance parameter problem in the quantum case. 
Otherwise, a claim of achievability for a bound might not be conclusive. 
Based on our general discussion, qubit models and several examples are examined to illustrate these findings. 

There are several issues that have not been explored in this paper. 
First is adaptive or sequential measurement schemes to implement an optimal measurement in the presence of nuisance parameters. 
In the subsequent paper, we continue to give a further analysis on one-parameter models with nuisance parameters. 
In particular, achievability of the SLD partial Fisher information shall be examined in detail. 

Second, the nuisance parameter problem is also important when estimating quantum channel parameters. 
For the channel estimation problem, we also has to optimize over input quantum states 
to extract as much information as possible. In the presence of nuisance parameters, 
an optimal input state also depends on them, and hence a similar tradeoff relation involves in general. 
We investigated the problem of detecting asymmetry in the qubit Pauli channel and 
showed that the effect of the nuisance parameter cannot be ignored \cite{gns19}. 

Last, practical efficient estimation strategies. 
In classical statistics, we do not intend to apply the best estimator when dealing with the nuisance parameter problem.  
But rather, we look for a sub-optimal estimator to suppress the effect of nuisance parameters. 
A typical example is to construct a pretty good estimator for the parameters of interest 
without any knowledge about the nuisance parameters. 
A challenge in the quantum case is to find a good measurement that does not depend 
on the nuisance parameters and to discuss how efficient we can extract information about the parameters of interest. 

The nuisance parameter problem in quantum systems is a common and practical problem 
when analyzing any statistical decision problem such as estimation, discrimination, control, and so on. 
Hence, it is relevant to any quantum information processing protocol in a noisy environment. 
As in the classical case, we need to develop proper statistical tools to handle them, 
since we cannot break the laws of statistics as well as quantum theory. 
This paper is just a beginning of research along this line, 
and we shall develop useful statistical tools in the subsequent publication.  

%%%%%%%%%%%%%%%%%%%%%%%%%%%%%%%%%%%%%%%%%
\section*{Acknowledgement}
The work is partly supported by JSPS KAKENHI Grant Number JP17K05571. 
The author is indebted to Prof.~ A.~Fujiwara and Prof.~H.~Nagaoka for invaluable discussions and suggestions. 
He would like to thank Dr.~K.~Yamagata for constructive comments on the work. 
He would also like to thank Prof.~H.~K.~Ng for discussions and her kind hospitality at Centre for Quantum Technologies 
in Singapore where part of this work was done. 

%%%%%%%%%%%%%%%%%%%%%%%%%%%%%%%%%%%%%%%%%%
\appendix %%%%%%%%%%%%%%%%%%%%%%%%%%%%%%%%%%%%%%%%%%
%%%%%%%%%%%%%%%%%%%%%%%%%%%%%%%%%%%%%%%%%%
\section{Logarithmic derivative operators and quantum Fisher information}\label{sec:AppQFI}
In this Appendix, we briefly summarize about the quantum score function and quantum Fisher information \cite{helstrom,holevo,petz}. 
We only consider the SLD Fisher information based on the symmetric logarithmic (SLD) operators. 
For a given smooth family of quantum states $\rho_{\theta}$ and any (bounded) linear operators $X,Y$ on $\cH$, 
Define the symmetric inner product by 
\[
\sldin{X}{Y}:=\frac12\tr{\rho_{\theta}(YX^\dagger+X^\dagger Y)},
\]
where $X^\dagger$ denotes the hermite conjugate of $X$. 
The $i$th SLD operator, $L_{\theta,i}$ is formally defined by the solution to the operator equation:
\be \label{sldrldop}
\frac{\del}{\del\theta_i}\rho_{\theta}=\frac12 (\rho_{\theta}L_{\theta,i}+L_{\theta,i}\rho_{\theta}), 
\ee
for $i=1,2,\dots,n$. 
The SLD Fisher information matrices are defined by 
\be\label{sldrld}
G_{\theta}:= \left[ \sldin{L_{\theta,i}}{L_{\theta,j}}\right]. 
%\tilde{G}_{\theta}:= \left[ \rldin{\tilde{L}_{\theta,i}}{\tilde{L}_{\theta,j}}\right], 
\ee
%respectively. 
It is convenient to introduce the following linear combinations of 
the logarithmic derivative operators. 
\[ 
L_{\theta}^i:= \sum_{j=1}^n({G}_\theta^{-1})_{ji}{L}_{\theta,j}.%,\ 
%\tilde{L}_{\theta}^i:=\sum_{j=1}^n(\tilde{G}_\theta^{-1})_{ji}\tilde{L}_{\theta,j} . 
\]
By definitions, $\{L_{\theta}^i\}$ forms a dual basis for 
the inner product space $\sldin{\cdot}{\cdot}$;  
$\sldin{L_{\theta}^i}{L_{\theta,j}}={\delta_{ij}}$.
%The same statement holds for the RLD case. 
The inverse of the SLD Fisher information matrice is expressed as
\[%\begin{align} \nonumber
G_\theta^{-1}=[g_\theta^{ij}]\mbox{ with }g_\theta^{ij}=\sldin{L_{\theta}^i}{L_{\theta}^j}.
\] %,\\ 

\section{Proofs}\label{sec:AppPr}
\subsection{Proof of Lemma \ref{lem_lucond}} \label{sec:AppPr1}
First, it is straightforward to see that the transformation \eqref{nui_change} 
preserves the first condition of Eq.~\eqref{lu_cond}. 
Note that the partial derivatives are transformed as 
\begin{align} \label{del_change1}
\frac{\del}{\del \xi_i}&= \frac{\del}{\del \theta_i}\quad(i=1,2,\dots,m),\\ \nonumber
\frac{\del}{\del \xi_j}&= \sum_{i=1}^n\frac{\del\theta_i}{\del\xi_j}\frac{\del}{\del \theta_i}\\ \label{del_change2}
&=\sum_{i=m+1}^n\frac{\del\theta_i}{\del\xi_j}\frac{\del}{\del \theta_i}\quad(j=m+1,\dots,n),
\end{align} 
where the relation $\del \xi_i/\del\theta_j={\delta_{ij}}$ for $i,j=1,2,\dots,m$ is used. 
Therefore, the condition $\frac{\del}{\del\theta_j}E_\theta[{\hat{\theta}_i}(X)|\Pi]=\delta_{ij}$ for $i,j=1,2,\dots,m$ 
remains same under the parameter transformation. 
Last, the second term of Eq.~\eqref{del_change2} together with condition \eqref{lu_cond} 
proves the relation 
$\frac{\del}{\del\xi_j}E_\xi[{\hat{\theta}_i}(X)|\Pi]=0$ for $i=1,2,\dots,m$ and $j=m+1,m+2,\dots,n$. 

\subsection{Derivation of expression \eqref{MICRbound2}}\label{sec:AppPr2}
Let us define another bound by
\[
C^{\Pi}_{\theta_I}[W_I,\cM_n]:=\min_{\Pi\mathrm{: POVM}} \Tr{W_I J^{\theta_I\theta_I}[\Pi] },
\] 
then, we will prove $C^{\Pi}_{\theta_I}[W_I,\cM_n]=C_{\theta_I}[W_I,\cM_n]$. 
The proof here is almost same line of argument as Ref.~\cite{nagaoka89}. 
Using the CR inequality for any locally unbiased estimator for $\theta_I$, 
we have 
\begin{align*}
C_{\theta_I}[W_I,\cM_n]&= \min_{\hat{\Pi}_I\mathrm{\,:l.u.\,for\,}\theta_I} \Tr{W_IV_{\theta_I}[\hat{\Pi}_I]}\\
&\ge \Tr{W_I J^{\theta_I\theta_I}[\Pi] }
\end{align*}
This is true for all POVMs that belong to the set . Therefore, we the relation $C_{\theta_I}[W_I,\cM_n]\ge C^{\Pi}_{\theta_I}[W_I,\cM_n]$. 
To prove the other direction, note that we can always construct a locally unbiased estimator 
at $\theta_0=\big(\theta_1(0),\dots,\theta_n(0)\big)$ for a given POVM $\Pi$. 
For example, 
\be\label{lu_est}
\hat{\theta}_i(X)= \theta_i(0)+\sum_{j=1}^n \left(J_{\theta_0}[e]\right)^{-1}_{ji}  \left.\frac{\del \log p_\theta(X|\Pi)}{\del\theta_j}\right|_{\theta_0}. 
\ee
Since this estimator is also locally unbiased about the parameter of interest $\theta_I$, 
and the MSE matrix about $\hat{\Pi}=(\Pi,\hat{\theta})$ satisfies $V_{\theta}[\hat{\Pi }]=J_\theta^{-1}[\Pi]$. 
In turn, we have a relationship $V_{\theta_I}[\hat{\Pi}_I]=J^{\theta_I\theta_I}[\Pi]$ for the parameter of interest. 
Thus, we obtain $C_{\theta_I}[W_I,\cM_n]\le C^{\Pi}_{\theta_I}[W_I,\cM_n]$. 
This proves $C_{\theta_I}[W_I,\cM_n]= C^{\Pi}_{\theta_I}[W_I,\cM_n]$. 

\subsection{Proof of Theorem \ref{thm_equiv}} \label{sec:AppPr3}
We shall prove the equivalence:
\[
C_{\theta_I}[W_I]= C^{\cE_\theta}_{\theta_I}[W_I]= C^{\lim}_{\theta_I}[W_I],
\]
holds for the bounds, 
\begin{align*}
C_{\theta_I}[W_I]
&=\min_{\hat{\Pi}_I\mathrm{:\,l.u.\,for\,}\theta_I}\Tr{W_IV_{\theta_I}[\hat{\Pi}_I]},\\
C^{\cE_\theta}_{\theta_I}[W_I]
&=\inf_{\hat{\Pi}_I\mathrm{:\,l.u.\,for\,}\theta}\Tr{W_IV_{\theta_I}[\hat{\Pi}_I]},\\
C^{\lim}_{\theta_I}[W_I]&=\lim_{W_n\to W_I}  \min_{\hat{\Pi}\mathrm{:\,l.u.\,for\,}\theta}\Tr{W_nV_{\theta}[\hat{\Pi}]}. 
\end{align*}
In this section, we omit the dependence of models in the above bounds for simplicity. 

We define two classes of POVMs as follows. 
Denote ${\mathscr M}_\theta$ by the set of POVMs $\Pi$ whose classical model 
$\cM(\Pi)=\{p_\theta(\cdot|\Pi)|\theta\in\Theta\}$ 
is regular, in particular, all score functions 
\be\nonumber
\left\{\frac{\del}{\del\theta_i} \log p_\theta(x|\Pi)\right\}_{i=1,2,\dots,n},
\ee
are linearly independent, and they span an $n$-dimensional tangent space at $\theta$. 
Let ${\mathscr M}_{\theta_I}$ by the set of POVMs $\Pi$ 
in which the linearly independent condition only holds for the parameter of interest, i.e., 
\be\nonumber
\left\{\frac{\del}{\del\theta_i} \log p_\theta(x|\Pi)\right\}_{i=1,2,\dots,m},
\ee
are linearly independent. 
Clearly, the inclusion ${\mathscr M}_{\theta}\subset{\mathscr M}_{\theta_I}\subset{\mathscr M}$ holds. 

First, let us prove the relation $C^{\cE_\theta}_{\theta_I}[W_I]= C_{\theta_I}[W_I]$. 
As mentioned in the main text, if an estimator is locally unbiased for all parameters, 
it is also locally unbiased for the parameter of interest. 
Therefore, we have 
\[
C^{\cE_\theta}_{\theta_I}[W_I]\ge C_{\theta_I}[W_I] .
\]
Next, let $\Pi^*=\arg\min_{\Pi\in{\mathscr M}_{\theta_I}} \Tr{W_IJ^{\theta_I\theta_I}[\Pi] }$ 
be an optimal POVM attaining the minimum. 
Since the resulting probability distribution $p^*_\theta(x)=\tr{\rho_\theta \Pi^*_{x}}$ 
may not be regular, one cannot construct a locally unbiased estimator about $\theta=(\theta_I,\theta_N)$ at $\theta$. 
In this case, we consider a randomized POVM $\Pi(\epsilon)$ as follows. 
Given $1>\epsilon>0$, we perform the POVM $\Pi^*$ with a probability $1-\epsilon$ 
and perform another POVM $\Pi^0\in{\mathscr M}_{\theta}$ with a probability $\epsilon$. 
The Fisher information matrix about this POVM is 
\[
J_\theta[\Pi(\epsilon)]=(1-\epsilon) J_\theta[\Pi^*]+\epsilon J_\theta[\Pi^0], 
\]
and this is regular. We can then find an locally unbiased estimator $\hat{\theta}_I$ for $\theta$ such that 
$\Tr{W_IV_{\theta_I}[\Pi(\epsilon),\hat{\theta}_I]}=\Tr{W_I J_\theta[\Pi(\epsilon)]^{-1}}$. 

Convexity of matrix inverse states 
\[
J_\theta[\Pi(\epsilon)]^{-1}\le (1-\epsilon) J_\theta[\Pi^*]^{-1}+\epsilon J_\theta[\Pi^0]^{-1}. 
\]
This then shows that
\[
\Tr{W_IJ_\theta[\Pi(\epsilon)]^{-1}} \le (1-\epsilon) C_{\theta_I}[W_I] +\epsilon \Tr{W_IJ_\theta[\Pi^0]^{-1}}. 
\]
Since $\Tr{W_IJ_\theta[\Pi^0]^{-1}}-C_{\theta_I}[W_I]\ge0$ for $W_I>0$, 
we find that 
\[
\Tr{W_IV_{\theta_I}[\Pi(\epsilon),\hat{\theta}_I]}\le C_{\theta_I}[W_I] +\delta, 
\]
holds for arbitrary $\delta>0$ by an appropriate choice of $\epsilon$ and $\Pi^0$. 
Therefore, $C_{\theta_I}[W_I] +\delta$ can be achieved by a locally unbiased estimator 
about $\theta$ to conclude 
$C^{\cE_\theta}_{\theta_I}[W_I]= C_{\theta_I}[W_I]$. 

The other relation $C^{\cE_\theta}_{\theta_I}[W_I]= C^{\lim}_{\theta_I}[W_I]$ is 
proven as follows. 
Rewrite the bound as $C_{\theta}[W_n]=\min_{\Pi\in {\mathscr M}_{\theta} } \Tr{W_n J_\theta[\Pi]^{-1}}$. 
Then, we have
\begin{align*}
C^{\lim}_{\theta_I}[W_I]&=\lim_{\epsilon\to0} C_{\theta}[W_n=W_I\oplus \epsilon I_N]\\
&=\lim_{\epsilon\to0} \min_{\Pi\in {\mathscr M}_{\theta} }
\left\{ \Tr{W_I J^{\theta_I\theta_I}[\Pi] }\right.\\
&\hspace{3cm}\left.+ \epsilon\Tr{J^{\theta_N\theta_N}[\Pi] }\right\}. 
\end{align*}
The term 
$\Tr{J^{\theta_N\theta_N}[\Pi]}$ is always positive for $\Pi\in{\mathscr M}_{\theta} $. 
Therefore, the following inequality holds. 
\begin{align*}
C^{\lim}_{\theta_I}[W_I]&\ge \lim_{\epsilon\to0} \inf_{\Pi\in {\mathscr M}_{\delta} }
 \Tr{W_I J^{\theta_I\theta_I}[\Pi] }\\
 &=C^{\cE_\theta}_{\theta_I}[W_I]. 
\end{align*}
Finally, we show that $C^{\lim}_{\theta_I}[W_I]+\delta$ 
can be achieved by a locally unbiased estimator about $\theta$ for arbitrary $\delta>0$. 
But this is clear from that for arbitrary $\epsilon$, 
we can always find a POVM $\Pi(\epsilon,\delta)$ such that 
$\Tr{W_I J^{\theta_I\theta_I}[\Pi(\epsilon,\delta)]}= C^{\lim}_{\theta_I}[W_I]+\delta$ holds. 
Since $\Pi(\epsilon,\delta)\in {\mathscr M}_{\theta}$, we can construct a locally unbiased estimator 
to conclude $C^{\cE_\theta}_{\theta_I}[W_I]= C^{\lim}_{\theta_I}[W_I]$.

\section{Classical CR inequality from a weight matrix optimization}\label{sec:AppSuppC}
In this section, we show that the classical CR inequality \ref{ccrineq2} in the presence of nuisance parameters 
can be derived from the proposed bound. In this section, we consider 
the optimization problem \eqref{qcrbound1_2}. 
Using Eq.~\eqref{qcrbound3}, the goal is to show 
\begin{multline}
\min_{\substack{W_N,W_{IN}=W_{NI}^{\mathrm T}:\\ W_n\ge0,W_I>0}}
\left[\Tr{W_n V_\theta} -\Tr{W_n J_\theta^{-1}}\right]\ge0\\
\Rightarrow V_{\theta_I} \ge J^{\theta_I\theta_I}. 
\end{multline}

The following lemma is well-known fact in matrix analysis (see, for example, Theorem 1.3.3 of Ref.~\cite{bhatia}).
\begin{lemma}\label{matrixineq}
For a $2\times 2$ block matrix on the complex number, 
\be
M=\left(\begin{array}{cc}A & B \\B^\dagger & D\end{array}\right), 
\ee
where $A\in\bbc^{m\times m}$, $B\in\bbc^{m\times n}$, and $D\in\bbc^{n\times n}$. 
Suppose $A>0$ and $D>0$, then $M\ge0$ holds if and only if $A\ge B D^{-1} B^\dagger$. 
\end{lemma}
It is easy to very that we can exchange the role of $A$ and $D$. 
We can also extend the above lemma for the case: $A>0$ and $D\ge0$. 
With this observation, we can rewrite the constraint for the minimization problem as 
$W_N,W_{IN}=W_{NI}^{\mathrm T}:\ W_N\ge W_{NI}W_I^{-1} W_{IN}$. 
Denote the difference of two matrices and its block-matrix representation 
according to the partition $\theta=(\theta_I,\theta_N)$ as
\begin{align}
M_\theta&:=V_\theta[\hat{\theta}]-J_\theta^{-1}\\ \nonumber 
M_\theta&=\left(\begin{array}{cc}M_{II} & M_{IN} \\M_{NI} & M_{NN}\end{array}\right),\ 
M_\theta^{-1}=\left(\begin{array}{cc}M^{II} & M^{IN} \\M^{NI} & M^{NN}\end{array}\right).  
\end{align}
Noting a matrix inequality $W_1\ge W_2$ implies $\Tr{V W_1}\ge \Tr{V W_2}$ for $V\ge0$, 
we have 
\begin{align*}
&\min_{W_n\ge0,W_I>0}\Tr{W_nM_\theta}\\
 \ge& 
\min_{W_N\ge W_{NI}W_I^{-1} W_{IN} }\Tr{W_I M_{II}} +\Tr{W_{IN} M_{NI}}\\
&\quad+\Tr{W_{NI}M_{IN}}+\Tr{W_{NI}W_I W_{IN} M_{NN}}\\
 =& 
\min_{W_N\ge W_{NI}W_I^{-1} W_{IN} }\Tr{W_I M_{II}} \\
&+\mathrm{Tr}\{(M_{NN}^{1/2}W_{NI}W_I^{-1/2} +M_{NN}^{-1/2}M_{NI}W_{I}^{1/2} )\\
&\times (M_{NN}^{1/2}W_{NI}W_I^{-1/2} +M_{NN}^{-1/2}M_{NI}W_{I}^{1/2} )^{\mathrm T}\}\\
&-\Tr{(M_{NN}^{-1/2}M_{NI}W_{I}^{1/2}) (M_{NN}^{-1/2}M_{NI}W_{I}^{1/2})^{\mathrm T}}\\ 
\ge&\Tr{W_I M_{II}}-\Tr{W_IM_{IN}M_{NN}^{-1}M_{NI}}\\
=&\Tr{W_I(M_{II}- M_{IN}M_{NN}^{-1}M_{NI})}. 
\end{align*}
Two inequalities can be saturated by the following 
choice of the weight matrix: 
\begin{align*}
W^*_{IN}&=-M_{NN}^{-1}M_{NI}W_I=(W^*_{NI})^{\mathrm T},\\
W^*_N&= W^*_{NI}W_I^{-1} W^*_{IN}
\end{align*}
Summarizing above argument, we prove
\begin{multline} \label{cwopt}
\min_{\substack{W_N,W_{IN}=W_{NI}^{\mathrm T}:\\ W_n\ge0,W_I>0}}\Tr{W_nM_\theta}\\=\Tr{W_I(M_{II}- M_{IN}M_{NN}^{-1}M_{NI})}. 
\end{multline}
The quantity in the right hand side is nonnegative, and thus we obtain
\begin{align*}
&\Tr{W_IM_{II}}\ge\Tr{ M_{IN}M_{NN}^{-1}M_{NI}}\ge0\\
&\Leftrightarrow\Tr{W_I V_{\theta_I}}\ge\Tr{W_I J^{\theta_I\theta_I}}. 
\end{align*}
Since this inequality holds for all positive matrices $W_I>0$, 
this is equivalent to the matrix inequality $V_{\theta_I} \ge J^{\theta_I\theta_I}$. 
This proves our claim. $\square$

We note that the above result immediately derives a tradeoff relationship among estimation errors in 
the classical parameter estimation problem. 
Expression \eqref{cwopt} shows that the matrix inequality 
\begin{multline}\label{ctradeoff} 
V_{\theta_I}[\hat{\theta}]\ge J^{\theta_I\theta_I}\\
+(V_{\theta_I\theta_N}-J^{\theta_I\theta_N}) (V_{\theta_N}-J^{\theta_N\theta_N})^{-1}
(V_{\theta_N\theta_I}-J^{\theta_N\theta_I}),
\end{multline}
holds for any locally unbiased estimator $\hat{\theta}$ at $\theta=(\theta_I,\theta_N)$. 
In this inequality, the second term of the right hand side, which is nonnegative, represents the tradeoff relationship. 
We can see that it vanishes if $V_{\theta_I\theta_N}[\hat{\theta}]-J^{\theta_I\theta_N}=0$. 
Lastly, we comment that this result itself is not surprising at all. 
Since we are imposing the locally unbiasedness condition for all parameters, 
We have the usual CR inequality $M_\theta[\hat{\theta}]=V_\theta[\hat{\theta}]- J_\theta^{-1}\ge0$. 
Using Lemma \ref{matrixineq}, this is equivalent to 
$M_{II}- M_{IN}M_{NN}^{-1}M_{NI}\ge0$. This is exactly the same statement as Eq.~\eqref{ctradeoff}. 

We point out that our weight-matrix optimization method can also provide the CR bound \ref{ccrineq1} 
without any nuisance parameter. Consider an $n$-parameter model and 
we wish to minimize the weighted trace of the inverse of the Fisher information matrix. 
This is expressed as $\min\Tr{W_nJ_\theta^{-1}}$ where the minimization 
is about $W_{IN}=W_{NI}^{\mathrm T}$ and $W_N$ under the constraint $W_n\ge$ and $W_I>0$. 
Working the exactly same procedure, we get
\begin{align*}
\min\Tr{W_nJ_\theta^{-1}}&=\Tr{W_I[J^{\theta_I\theta_I}- J^{\theta_I\theta_N} (J^{\theta_N\theta_N})^{-1}J^{\theta_N\theta_I}]}\\
&=\Tr{W_I J_{\theta_I\theta_I}^{-1}}.
\end{align*}
Here the second line is due to the Schur's complement. 
Therefore, the CR inequality $V_{\theta_I}[\hat{\theta}]\ge J_{\theta_I\theta_I}^{-1}$ holds 
by considering the CR type bound among all possible weight matrices for the nuisance parameter.


\begin{thebibliography}{99}
\bibitem{com_lu}
Unbiased estimators are often discussed in standard textbooks, but 
locally unbiased estimators are not so common among statisticians. 
We find the latter concept important when discussing the nuisance parameter problem 
in quantum estimation theory.
\bibitem{com_reg}
To avoid mathematical subtleties, we need to impose regularity 
conditions for quantum statistical models. For example, 
a mapping $\theta\mapsto \rho_\theta$ is one-to-one and smoothness 
so that we can differentiate $\rho_\theta$ sufficiently many times. 
$\del\rho_\theta/\del\theta_i$ are linearly independent. 
In the quantum case, we need to draw attention to 
the rank of quantum states. For the sake of clarity, 
we only consider full-rank states only in this paper. 
For problems in the pure-state model, see for example Refs.~\cite{fn95,fn99}.
\end{thebibliography}
\end{document}